\documentclass[sn-nature]{sn-jnl}% Style for submissions to Nature Portfolio journals
\usepackage{graphicx}%
\usepackage{multirow}%
\usepackage{amsmath,amssymb,amsfonts}%
\usepackage{amsthm}%
\usepackage{mathrsfs}%
\usepackage[title]{appendix}%
\usepackage{xcolor}%
\usepackage{textcomp}%
\usepackage{manyfoot}%
\usepackage{booktabs}%
\usepackage{algorithm}%
\usepackage{algorithmicx}%
\usepackage{algpseudocode}%
\usepackage{listings}%

\usepackage{hyperref}

\definecolor{listcolor}{RGB}{0, 0, 0} % Define the color you want
\hypersetup{
    colorlinks=true,
    linkcolor=black, % Set hyperlink color to black
    urlcolor=blue,
}
\usepackage{subfig}
\usepackage{colortbl}
\definecolor{LightCyan}{rgb}{0.88,1,1}

\begin{document} 

%\title{Beyond IRA: Exploring the Influence of Suspended Accounts in Twitter Discourse During the 2016 US Election.}
\title{{\color{blue} Suspended accounts align with the Internet Research Agency misinformation campaign to
influence the 2016 US election}}
%%=============================================================%%
%% Prefix	-> \pfx{Dr}
%% GivenName	-> \fnm{Joergen W.}
%% Particle	-> \spfx{van der} -> surname prefix
%% FamilyName	-> \sur{Ploeg}
%% Suffix	-> \sfx{IV}
%% NatureName	-> \tanm{Poet Laureate} -> Title after name
%% Degrees	-> \dgr{MSc, PhD}
%% \author*[1,2]{\pfx{Dr} \fnm{Joergen W.} \spfx{van der} \sur{Ploeg} \sfx{IV} \tanm{Poet Laureate} 
%%                 \dgr{MSc, PhD}}\email{iauthor@gmail.com}
%%=============================================================%%

 \author*[1,6]{\fnm{Matteo} \sur{Serafino}}\email{m.serafi00@ccny.cuny.edu}

 \author[2,6]{\fnm{Zhenkun} \sur{Zhou}}%\email{iauthor@gmail.com}

 \author[3]{\fnm{Jos\'e} \sur{S. Andrade, Jr.}}%\email{iiauthor@gmail.com}
% %\equalcont{These authors contributed equally to this work.}

 \author[4,5]{\fnm{Alexandre} \sur{Bovet}}%\email{iiiauthor@gmail.com}
% %\equalcont{These authors contributed equally to this work.} 

 \author*[1]{\fnm{Hern\'an} \sur{A. Makse}}\email{hmakse@ccny.cuny.edu}
% %\equalcont{These authors contributed equally to this work.}

 \affil[1]{\orgdiv{Levich Institute and Physics Departmen}, \orgname{City College of New York}, \orgaddress{\city{New York}, \state{NY}, \country{USA}}}

 \affil[2]{\orgdiv{Department of Data Science, School of Statistics}, \orgname{Capital University of Economics and Business}, \orgaddress{ \city{Beijing}, \country{China}}}

 \affil[3]{\orgdiv{Physics Department}, \orgname{Universidade Federal do Cear\'a}, \orgaddress{\city{Fortaleza}, \state{Cear\'a}, \country{Brazil}}}

 \affil[4]{\orgdiv{Department of Mathematical Modeling and Machine Learning}, \orgname{University of Zurich}, \orgaddress{\city{Zurich}, \country{Switzerland}}}

 \affil[5]{\orgdiv{Digital Society Initiative}, \orgname{University of Zurich}, \orgaddress{\city{Zurich}, \country{Switzerland}}}
\affil[6]{These authors contributed equally: Matteo Serafino, Zhenkun Zhou}

%%==================================%%
%% sample for unstructured abstract %%
%%==================================%%

\abstract{The ongoing debate surrounding the impact of the Internet Research Agency's (IRA) social media campaign during the 2016 U.S. presidential election has largely overshadowed the involvement of other actors. Our analysis brings to light a substantial group of suspended Twitter users, outnumbering the IRA user group by a factor of 60, who align with the ideologies of the IRA campaign. Our study demonstrates that this group of suspended Twitter accounts significantly influenced individuals categorized as undecided or weak supporters, potentially with the aim of swaying their opinions, as indicated by Granger causality.}


%%\pacs[JEL Classification]{D8, H51}

%%\pacs[MSC Classification]{35A01, 65L10, 65L12, 65L20, 65L70}

\maketitle
\clearpage
\newpage
\section{Introduction}\label{sec1}
Social media platforms have become increasingly prominent in shaping
political events and social discussions. Political campaigns across
the globe are heavily reliant on social media platforms to communicate
with the masses and shape public opinion \cite{digrazia2013more, 
anstead2015social,bovet2018validation, ahmed20162014, majo2021role}. 
However, the rise of social media has also resulted in debates about 
their impact on society and the potential risks associated with their use.

Social media platforms, while holding the potential to facilitate 
communication and foster informed discussions, are also susceptible 
to the dissemination of misinformation and disinformation campaigns 
\cite{hegelich2016social, ratkiewicz2011detecting,bruno2022brexit}. This issue extends
beyond politics and seeps into sensitive domains like public health,
as exemplified by the anti-vaccine movements during the COVID-19 
pandemic \cite{burki2020online}. Compelling evidence abounds, pointing
to the active exploitation of social media platforms by certain governments
to subvert domestic social movements and interfere in the democratic
elections of foreign adversaries \cite{tucker2017liberation}. Noteworthy
instances of such foreign interventions include the case of the 2017 
French presidential election \cite{ferrara2017disinformation} and 
the highly significant interference by the Internet Research Agency
(IRA: a Russian company engaged in online influence operations on 
behalf of Russian business and political interests) in the 2016 US
presidential election \cite{TheDisinformationReport, jamieson2020cyberwar}.

As outlined in the U.S. Special Counsel's report~\cite{mueller2019report}, 
the Internet Research Agency initiated Russian interference operations
as early as 2009. Their strategic approach involved the creation of social
network campaigns aimed at fueling and magnifying political and social
divisions within the United States \cite{mueller2019report,carroll2017st}. 
At the beginning of 2018, Twitter committed to the United States Congress and 
the public to provide regular updates and information regarding their investigation 
into foreign interference in U.S. political conversations on Twitter. In October
2018, Twitter openly released all the accounts and related content associated with
potential information operations they had found on Twitter since 2016. This 
dataset consists of more than three thousand accounts affiliated with the IRA. 
It contains more than 9 million tweets, including the earliest Twitter activity
of the accounts connected with these campaigns, dating back to 2009. The Twitter
corporation estimates that 9\% of the tweets from IRA accounts were election-related.

Since then, the number of works focusing on the role the IRA agency played in 
the 2016 US political campaign and social debates increased. A. Badawy {\it et al.}\cite{badawy2018analyzing} found that conservatives retweeted Russian trolls 
significantly more often than liberals and produced 36 times more tweets. 
Among the 5.7 million distinct users analyzed between September 16 and November 9,
2016, about 4.9\% and 6.2\% of liberal and conservative users, respectively,
were automated accounts (bots) used to share troll content. Text analysis of the
content shared by trolls reveals that they had a mostly conservative, pro-Trump agenda. 
P. N. Howard {\it et al.}  \cite{howard2018ira} concluded that the Russian
strategies targeted many communities within
the United States, particularly the most extreme conservatives and those with 
particular sensitivities to race and immigration. They found that IRA used a variety
of fake accounts to infiltrate political discussions in liberal and conservative
communities, including black activist communities, to exacerbate social 
divisions and influence the agenda. By combining network science and volumetric analysis, L. G. Stewart
{\it et al.} found that troll accounts shared content to polarized information
networks, likely accentuating disagreement and fostering division \cite{stewart2018examining}.
The conclusions above align with the findings of R. DiResta {\it et al.} 
\cite{diresta2018tactics}, who observed that the IRA campaign was designed to exploit societal fractures,
blur the lines between reality and fiction, and erode trust in media entities and the information
environment, in government, in each other, and in democracy itself. In their study 
on disinformation S. Zannettou {\it et al.} \cite{zannettou2019disinformation}, conducted an 
investigation into the behavioral differences between IRA and random Twitter users. The findings 
revealed that IRA users exhibit a higher tendency to disseminate content related to politics.
Additionally, IRA employed multiple identities throughout the lifespan  of their accounts and 
made deliberate efforts to amplify their impact on Twitter by increasing their number of followers.

The studies mentioned above aim to characterize the IRA campaign. 
In an attempt to evaluate the impact of the IRA campaign on Twitter users, 
C. Bail {\it et al.} conducted a study using a longitudinal survey that 
describes the attitudes and online behaviors of one thousand Republican
and Democratic Twitter users in late 2017 \cite{bail2020assessing}. Their findings
suggest that Russian trolls might have failed to sow discord because they mostly
interacted with those who were already highly polarized.
In \cite{grinberg2019fake}, N. Grinberg {\it et al.} demonstrated that exposure to fake
news content during the 2016 elections was typically concentrated among a small group
of users, particularly those who identify themselves as strong political partisans. If exposure 
to social media posts from Russian foreign influence accounts during the 2016 US election was 
similarly concentrated, their impact on changing candidate preferences may have
been minimal. In the attempt to verify this hypothesis, Eady {\it et al.} \cite{eady2023exposure}
combined US longitudinal survey data from over 1496 respondents with Twitter data. 
They found that exposure to the Russian foreign influence campaign was heavily
concentrated among a small fraction of users who identified themselves as Republicans. 
Moreover, they found no evidence of a significant relationship between exposure
to the campaign and changes in attitudes, polarization, or voting behavior in the 
2016 US election.

While prior research extensively
explored the influence of IRA accounts on individuals' voting intentions
and Twitter discussions, it provided limited attention to
the broader set of suspended accounts not flagged as IRA.
During the "Twitter purge" in May 2018, Twitter suspended numerous
accounts, including those unrelated to the
IRA \cite{bursztein2018quantifying, roth2018twitter}, with
IRA accounts constituting only a smaller fraction of this overall set.

Considering that all these accounts faced suspension for violating
Twitter rules, it prompts the question of whether, beyond IRA accounts,
other accounts might have attempted to influence Twitter discourse and
more broadly the 2016 US election.

Our findings provide evidence supporting this notion. A consistent group of
suspended accounts exhibits similarity with IRA accounts in terms of
the information they interact with and disseminate to the broader Twitter community.
{\color{blue}{We demonstrate that the group of suspended accounts did indeed influence, in a Granger-causal manner, the retweet activity of undecided users and weak supporters—individuals uncertain about their voting decisions—in terms of political polarization.}}

The paper is organized as follows: Section \ref{methods} outlines the data
collection and analysis methods utilized throughout the study. In Section
\ref{Section1}, we present the findings of our research, which include:
{\bf a)} the characterization of users within our dataset based on the
content they share; {\bf b)} the use of the IRA ego network as a means
to identify a group of suspended users with similar behavioral patterns;
and {\bf c)} an evaluation of how this specific group of suspended accounts
influences Twitter discourse, utilizing Granger causality for assessment. 
The manuscript concludes with a thorough discussion and conclusion section,
summarizing the key insights gained from our analysis.

\section{Methods}
\label{methods}
\subsection{Dataset}
\label{data}

In this study, we combine the IRA dataset with a dataset containing tweets
posted between June 1st and election day, November 8th, 2016. The data
were collected continuously using the Twitter search API with the names 
of the two presidential candidates \cite{bovet2018validation,bovet2019influence, 
flamino2023political}. The 2016 dataset consists of 171 million tweets
sent by 11 million users. 

On the other hand, from June 1st to November 8th, 2016, 556 IRA accounts published
391680 tweets in English. According to \cite{howard2018ira}, the content of these
tweets aimed to sow and amplify political and social discord in the United States
and manipulate the 2016 American presidential election. See Supplementary Sections 1 and 2
for more information.

To retrieve the account status of each user in the 2016 dataset,
we used the Twitter users API, as of October 2023. It allows us to classify each
account as suspended, not found, not verified, or verified. On Twitter,
a suspended account refers to an account that has been temporarily or
permanently disabled by Twitter due to a violation 
of its rules or policies. In contrast, a not found account is not deleted by Twitter 
but is no longer available because the user has chosen to delete or deactivate 
it. A not verified account on Twitter is an account that has not been officially
confirmed by Twitter. Verification is a process through which Twitter 
verifies the authenticity and identity of notable public figures, organizations,
or brands. On the other hand, a verified account on Twitter has undergone the 
verification process and has been confirmed by Twitter as an authentic
representation of a notable public figure, organization, or brand. Verified 
accounts are distinguished by a blue checkmark badge next to their username,
indicating their credibility and authenticity. Important to note that this
is no longer the case (as of November 29th, 2023), as now anyone can buy the
blue checkmark. 

Among the 11 million users, 73.8\% are not verified, 17.7\% are not found,
7.7\% are suspended, and 0.8\% are verified. In this dataset, the IRA accounts
account for less than 1\% of the users (554 accounts). 

\subsection{News categories}
\label{categories}

In order to control for the type of information under analysis, we focus on
tweets that contain at least one URL (Uniform Resource Locator) pointing
to a news website outside of Twitter. We classified URL links for outlets
that mostly conform to professional standards of fact-based journalism in
five news media categories: right, right leaning, center, left leaning, and left. 
The  classifications rely on the website \href{https://www.allsides.com/unbiased-balanced-news}{allsides.com} 
(AS), followed by the bias classification from the website \href{https://mediabiasfactcheck.com/}
{mediabiasfactcheck.com} (MBFC) for outlets not listed in AS (both accessed on 7 January
2021 for the 2020 classification) \cite{bovet2018validation, bovet2019influence,flamino2023political,
grinberg2019fake}. We also include three additional news media categories to 
include outlets that tend to disseminate disinformation: Extreme bias right,
Extreme bias left, and Fake news \cite{bovet2018validation,bovet2019influence,flamino2023political,
grinberg2019fake}. Websites in the fake news category have been flagged by 
fact-checking organizations as spreading fabricated news or conspiracy theories, 
while websites in the extremely biased categories have been flagged for reporting 
controversial information that distorts facts and may rely on propaganda,
decontextualized information or opinions misrepresented as facts. Supplementary 
Table 1 offers the list of news outlets per category considered in this work.

In the 2016 dataset, 2.3 million users shared 30.7 million tweets that contained 
URLs directing to news outlets. In the IRA dataset, 334 IRA accounts posted 23,806
tweets that included hyperlinks to news outlets.  

\subsection{Retweet network}
\label{retweet}
In the context of a news category network, a link between two users occurs
every time a user $u$ retweets a user's tweet $v$ 
that contains a URL linking to a website belonging to one of the news media categories.
The direction of the connection goes from $v$ to $u$, i.e., the direction of the
information flow between Twitter users. We do not include multiple links in the same 
direction between the same two users, nor do we include self-links. The degree of a node
in the network is defined as the number of edges connected to it. The out-degree of a 
node $u$, $ k_{out}^u $, represents the number of unique users who retweeted $u$. On the other
hand, the in-degree of a node $u$, $ k_{in}^u$, represents the number of users retweeted
by node $u$. It is worth noticing that, by construction, these networks are 
balanced directed networks, and as such, $ \langle k_{in} \rangle = \langle k_{out} \rangle =
\langle k \rangle /2$. 

When building the ego networks, we proceed similarly. 
We construct separate networks for each of the four types of 
interactions: retweeting, mentioning, replying, and quoting. Each node in 
the network represents either an IRA or a non-IRA user. A link between two users 
occurs every time a user $u$ interacts with a user $v$ through a type of interaction.
The direction of the connection goes from $v$ to $u$, i.e., the
direction of the information flow. Connections are allowed between IRA nodes 
and between IRA and non-IRA nodes. We do not consider interactions among non-IRA nodes. 
We consider multiple interactions between two users; that is, networks are weighted by 
the number of times users interact.

Starting from the four interaction networks, we build an aggregated
network (referred to as IRA ego network or IRA aggregated ego network,
interchangeably) by considering all types of interaction and removing self-loops.
As usual, an edge connecting node $u$ with node $v$  means there was at least one
type of interaction between them. The edge is 
weighted by the number of interactions among $u$ and $v$. The directions of
the links are according to the flow of information. 
The resulting structure is a directed weighted network of 179,783 nodes 
(524 of which are IRA accounts) and  432,429 edges (see Table~\ref{table:four}).

\subsection{Sampling strategies}
\label{Sampling}

To avoid sample bias, we randomly extracted the same amount
as the number of IRA users for each group and category
(making sure not to select the IRA users). We average the in/out degree
over 1000 realizations. For each realization, sampling 
was without replacement. We refer to them as $(\overrightarrow{ k_{type}^s} ^i, 
\overrightarrow{ k_{type}^s } ^j)$, where the superscript $s$ indicates that the 
degree considered comes from the sampled nodes. Supplementary Table 7 
displays the sampled average degree for each
group in each news category, together with the standard error. 
Refer to Supplementary Table 8 for a view of the non-sampled case. 

\subsection{Two-sample Kolmogorov-Smirnov test}
\label{KS}
To test for differences in the in/out-degree activity of suspended, not found,
not verified, verified, and IRA accounts, we employed a Two-sample Kolmogorov-Smirnov
test with null hypothesis $H_0$: $F_i(x) = F_j(x)$ where $x =  k_{type}^s$. 
The superscript $s$ indicates that the degree considered comes from the 
sampled nodes, $i,j \in $  (suspended, not found, verified, not verified, IRA), and $type \in$ (in, out). 
The null hypothesis, denoted as $H_0$,
assumes that the activity of a given user in each interaction type, 
represented by $x = (x_{out},x_{in})$, follows the condition $F_{out}(x) =
F_{in}(x)$ for every $x$. Here, $F_{out}(x)$ and $F_{in}(x)$ represent
the cumulative density functions (CDF) for the ``out'' and ``in'' directions,
respectively. The alternative hypothesis, on the other hand, suggests
that $F_{out}(x) <  F_{in}(x)$  (or $F_{out}(x) >  F_{in}(x)$) for at least one $x$.

It is worth noting that these hypotheses describe the CDFs of the underlying 
distributions, not the observed data values. For example, suppose $x_{out} 
\sim F_{out}$ and $x_{in} \sim F_{in}$. If $F_{out}(x) > F_{in}(x)$ for
all $x$, the values $x_{out}$ tend  to be less than $x_{in}$.
We set a level of 5\%, meaning that we will reject the null hypothesis
and favor the alternative if the p-value is less than 0.05.

\subsection{Supporters identification}
\label{supporters}

Similarly to \cite{bovet2018validation}, we use a supervised classifier to classify each
tweet in favor of Donald Trump or Hillary Clinton. The training set was built using the
hashtag co-occurrences network to investigate Twitter users' opinions on the two presidential
candidates. We classified a user as a supporter of Trump if the number of her/his tweets 
supporting  Trump $N_{\textit{pro-T}}$ is greater than the number of tweets supporting Clinton
$N_{\textit{pro-C}}$. We define the support of a given user toward the candidates 
as $S=N_{\textit{pro-T}}- N_{\textit{pro-C}}$. If $S>0$, the user supports Trump. Otherwise, 
the user is likely to support Hillary. The highest the value of $S$ in absolute terms, the 
strongest the support. Considering all the users in the dataset, 65\% of them support Hillary 
Clinton while 28\% are in favor of Donald Trump (7\% are unclassified as they have the same
number of tweets in each camp)~\cite{bovet2018validation}. When considering only the users
interacting with IRA accounts, 25\% of the users are classified as Clinton supporters,
and 72.6\% of the users are classified as Trump supporters. 

\subsection{Supporting classes}
\label{classes}

To distinguish between strong and weak supporters based on their $S$ values, we utilize 
the interquartile range (IQR) of $S$, defined as $IQR = Q_3 - Q_1$, where $Q_3$ represents
the third quartile and $Q_1$ represents the first quartile. In this analysis, a positive value of
$S$ indicates a likelihood of supporting Trump, while a negative value of $S$ suggests a preference for 
Clinton. The magnitude of $S$ quantifies the degree of support for a particular candidate.
Users who consistently retweet in favor of Trump referred to as strong supporters, exhibit 
higher $S$ values. Conversely, users with significantly negative $S$ values can be associated 
with strong supporters of Clinton. Users with $S=0$ are categorized as undecided since they 
display an equal number of tweets supporting both candidates.

In addition to the undecided category, we define four classes of supporters based on the
interquartile range (IQR) of $S$ values. For Trump supporters, the IQR is calculated over 
the values of $S>0$, while for Clinton supporters, the IQR is computed using the absolute
values of $S<0$. We identify weak Trump (Clinton) supporters as users whose $S$ values 
fall below $Q_3 + 1.5IQR$. On the other hand, 
strong Trump (Clinton) supporters are individuals whose $S$ values exceed $Q_3 + 1.5IQR$. 
This classification scheme allows us to distinguish between different levels of support.

Alternatively, we can consider the entire distribution of $S$ and define strong Trump
supporters as users with $S$ values above $Q_3 + 1.5IQR$. Similarly, strong Clinton 
supporters are those with $S$ values below $Q_1 - 1.5IQR$. Weak Trump supporters fall 
within the range $Q_1 - 1.5IQR \leq S \leq Q_3 + 1.5IQR$ with $S>0$, while weak Clinton
supporters fall within the same range but with $S<0$. This alternative classification,
in the case of users interacting with IRA, results in 12.7\% of users identified as strong
Trump supporters, 60\% as weak Trump supporters, 2.6\% as strong Clinton supporters, and 
22.5\% as weak Clinton supporters. These percentages slightly differ from the ones 
obtained using the other approach mentioned in the main paper.

\section{Results}
\label{Section1}

\subsection{Accounts characterization}
\label{characterization}
Our analysis commences with a general characterization of the accounts
active on the Twitter platform around the topic ``election" during the
2016 US presidential elections. Refer to the Methods section, 
specifically Section \ref{data} for a comprehensive description of the dataset.
Users in this dataset are classified into distinct groups, encompassing
IRA-flagged accounts by Twitter, along with not found, not verified,
verified, and suspended accounts. This section aims to characterize
the various account groups in terms of the information they spread on
Twitter and draw comparisons with the IRA-flagged accounts. The rationale
for using IRA as a benchmark in our analyses, as explained in the introduction,
is to assess the impact of other accounts displaying behavior similar to IRA,
specifically those favoring the right political candidate.

\begin{figure}[!ht]
  \centering
  \subfloat{{\bf (a)}\includegraphics[width=0.4\linewidth]{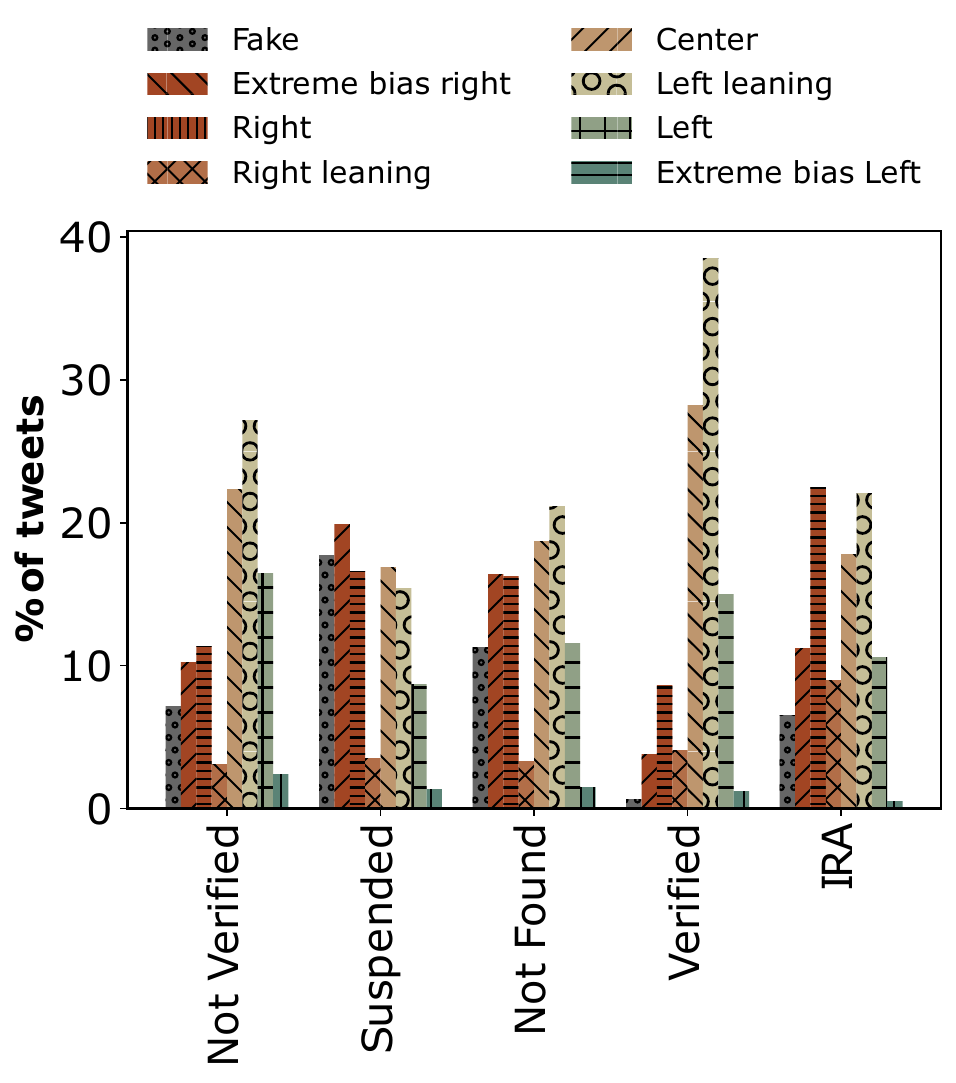}}
  \subfloat{{\bf (b)}\includegraphics[width=0.4\linewidth]{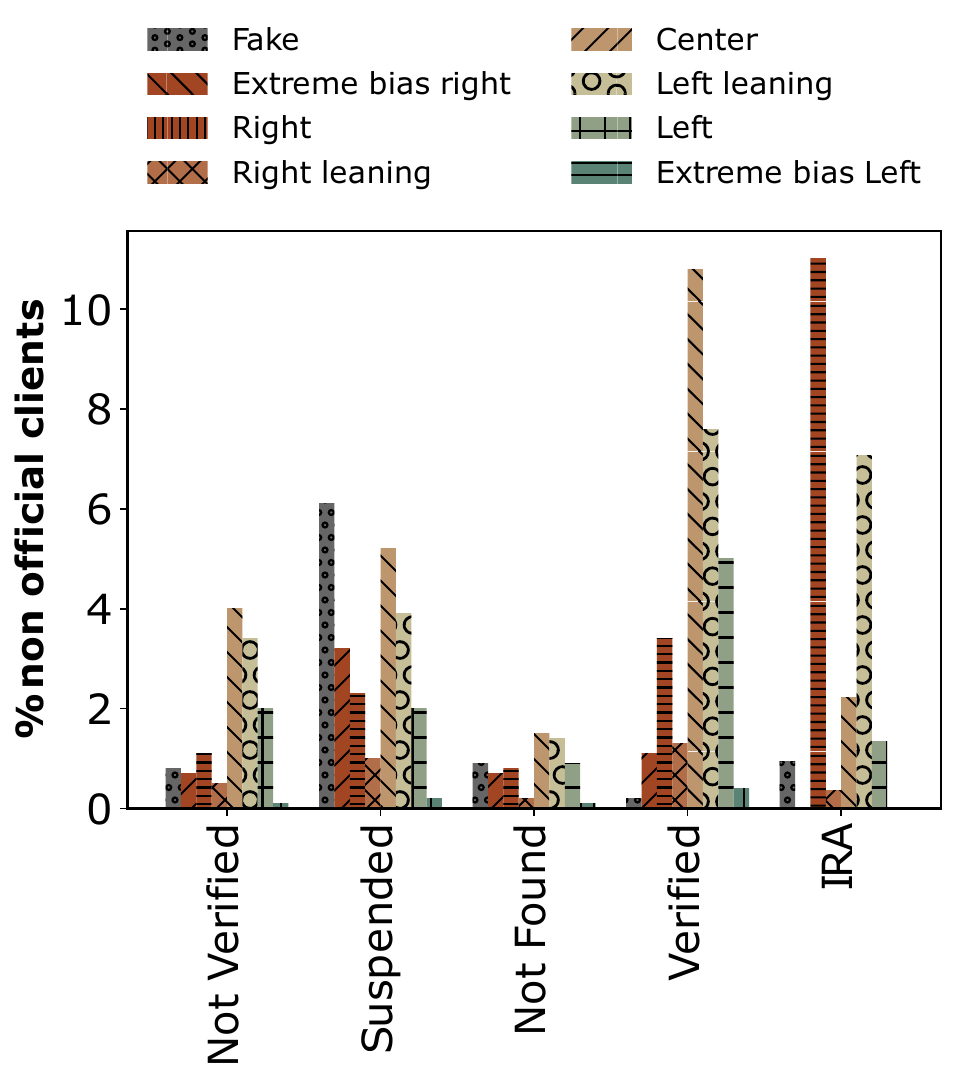}}\\
  \caption{{\bf  Distribution of tweets and clients' type per account type.}
   {\bf (a)} The fraction of tweets with a URL pointing to a 
  website belonging to one of the categories. Normalization is computed per
  group, meaning that, for example, the fraction of not verified
  tweets per category sums up to one. In each of the five groups, the order 
  of bars is kept the same. We always display bars in the following orders: fake news, 
  extreme bias right, right, right leaning, center, left-leaning, left, 
  and extreme bias left. See Supplementary Table 2 for further details.
  {\bf (b)} The fraction of tweets with
  a URL pointing to a website belonging to one of the categories posted
  through non official sources. Refer to Supplementary Table 3 for a complete
  view of the percentages. }
 \label{fig:1}
\end{figure}

The initial distinction among account types concerns the kind of information
users engage with (see Section \ref{categories}). 
Verified accounts, as shown in Fig.~\ref{fig:1}a, have a
higher fraction of center and left-related (left and left-leaning) tweets,
while IRA, suspended, and not found accounts exhibit a higher fraction of
right-related tweets. Unlike IRA accounts, which show a significant percentage
of center and left-leaning related news, suspended accounts have the lowest
fraction of left-related content and the highest fraction of fake-related content.
Supplementary Tables 2 and 3 provide a full breakdown of these percentages.

In Fig.~\ref{fig:1}b, we show the percentage of tweets shared
through non official clients for each media category. To ensure comparability,
we normalize the percentages per account type by the total group activity,
including both official and non official clients. 
For official client details, refer to Supplementary Table 5.
Analyzing tweet clients offers valuable insights into tweet origins,
especially their potential bot-generated nature. Non official clients,
encompassing applications like \href{https://ifttt.com/}{ifttt} and
\href{https://dlvrit.com/}{dlvrit}, span professional automation
tools to manually programmed bots.

Figure~\ref{fig:1}b shows that verified accounts exhibit the
highest fraction of tweets from non official clients, 
constituting 22.9\% of their total activity. The
most frequently used clients for verified accounts, such as
\href{https://www.hootsuite.com/}{Hootsuite} and
\href{https://piano.io/it/product/socialflow/}{Socialflow},
are renowned for automating interactions within the Twitter ecosystem. 
These verified accounts, often belonging to journalists or public figures,
utilize such tools for social media activities. Suspended accounts rank second
at 23.9\%, with \href{https://dlvrit.com/}{Dlvrit} as their primary client,
closely followed by IRA accounts at 22.9\%, mostly relying on
\href{https://twitterfeed.com/}{Twitterfeed} for automated activity. Not verified
and not found accounts exhibit below 13\% non-official client usage,
with \href{https://twitterfeed.com/}{Twitterfeed} being the most used client.
It is noteworthy that verified accounts predominantly used non official clients
to disseminate center, left-leaning, and left news. In contrast, IRA accounts
utilized non official clients mainly for right-related content, with a smaller
percentage of left-leaning material. Suspended accounts employed non-official
clients primarily for fake news dissemination.

\begin{figure}[ht!]
  \centering
  \includegraphics[width=0.40\linewidth]{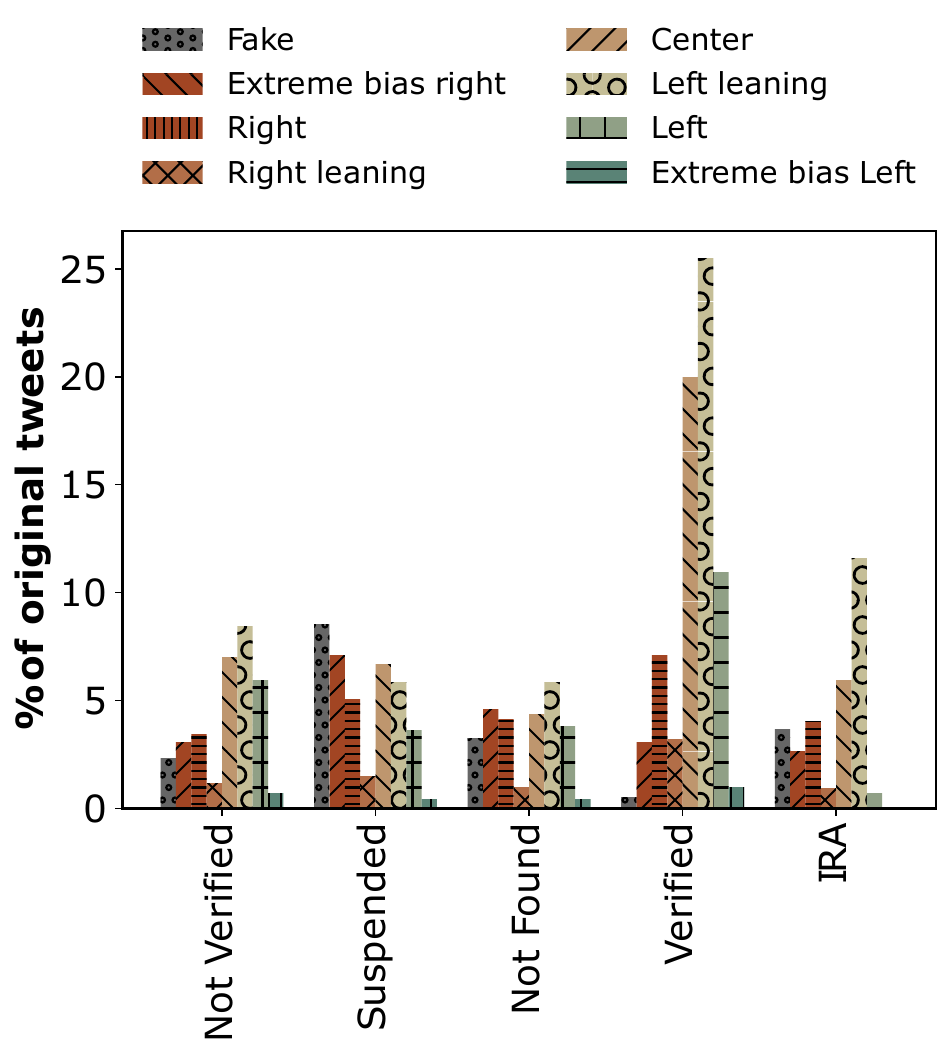}
  \caption{{\bf Share of original tweets per group.} The percentage of
  original tweets per group and news category is presented, with normalization
  conducted over the total activity of each group. This means that
  the sum of the percentages per group represents the total fraction 
  of original tweets. Further details can be found in Supplementary Table 6.}
 \label{fig:5}
\end{figure}

Figure~\ref{fig:5} shows the proportion of original tweets shared by each group 
and category. The normalization is over the total activity of each account type, 
meaning that the sum of the percentages per each account type represents the total 
fraction of original tweets. The group with the highest share of original tweets
is the verified one, with a value of 71.2\%. This group also shows the lowest share 
of original tweets linking to fake news and the highest share of original tweets  
related to the center, left-leaning, and left categories. IRA accounts instead
show together with not found accounts, the lowest share of original 
tweets, with a percentage of 29.4\% and 27.3\%, respectively. Most of the original tweets
shared by IRA belong either to the left leaning or center categories.
Not found accounts have a more homogeneous distribution of original tweets among the 
different categories. Suspended accounts, with 38.6\% of original tweets, show the highest
percentage of original tweets related to the fake category and the extreme 
bias right category. Not verified accounts (32\% of original tweets) 
show higher percentages in the center and left leaning.

Finally, we test whether suspended, not found, verified, not verified,
and IRA behave differently in terms of their in/out activity in each
news category network (see Section \ref{retweet} and Table~\ref{table:one}).
We employ a two-sample Kolmogorov-Smirnov test
\cite{hodges1958significance,mann1947test} (two-sided version, 
see Section \ref{KS}) with null hypothesis $H_0$: the data are drawn
from the same distribution. We performed the test for each two-pair
combination of the groups $(\overrightarrow{ k_{type} } ^i,
\overrightarrow{ k_{type} } ^j)$, with $i,j \in $ 
(suspended, not found, verified, not verified, IRA), and $type \in$ (in, out). 
The $\overrightarrow{ k_{type}}$ vector contains the
values of $\langle k_{type} \rangle$ for each category network.
We adapt sampling strategies to avoid sampling bias error, as explained
in Section \ref{Sampling}. Refer to Supplementary Table 7 for a view
of the sample degrees. {\color{blue} While the news category networks considered
here are not weighted, we did check for differences in the weighted case
and could not find any significant variations.}

\begin{figure}[ht!]
  \centering
  \includegraphics[width=0.40\linewidth]{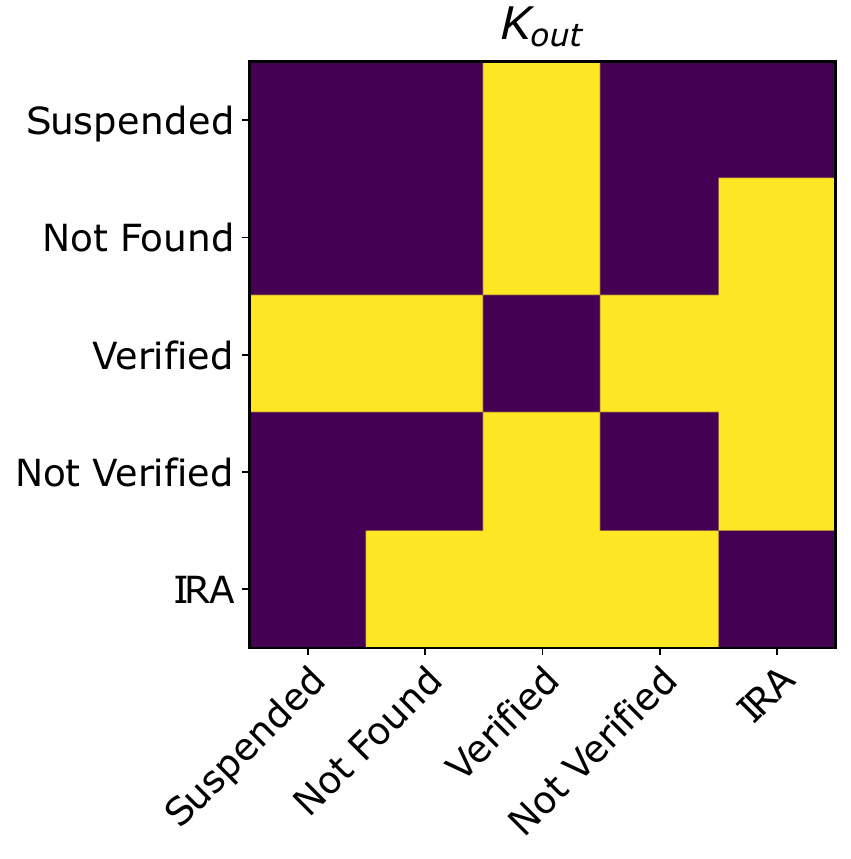}
  %\subfloat{{\bf (b)}\includegraphics[width=0.40\linewidth]{Fig4b.pdf}}
  \caption{{\bf Two-sample Kolmogorov-Smirnov tests.} Results of the two sample
  Kolmogorov Smirnov test between groups based on their out degree.
  {\color{blue}Yellow boxes indicate rejection of the null
  hypothesis in favor of the two-sided alternative, suggesting that the data were
  not drawn from the same distribution.} We observe that $H_0$ 
  is only rejected between IRA and suspended accounts, indicating a similarity in
  the out-degree activity of these two groups. Furthermore, verified accounts
  exhibit an out-degree activity that differs from that of the other groups.}
 \label{fig:4}
\end{figure}

Figure~\ref{fig:4} shows the results of the tests for the
out-degree and in-degree, respectively. We used a heatmap representation,
where the yellow color indicates the rejection of the null hypothesis $H_0$ 
in favor of the default two-sided alternative, suggesting that the data were
not drawn from the same distribution.

In comparing out-degree activity, verified accounts consistently exhibit
distinct behavior from other groups, rejecting the null hypothesis. 
IRA accounts, however, display similar behavior to suspended accounts
while differing from verified, not found, and not verified accounts. 
In terms of in-degree results, verified accounts differ from suspended, 
not found, and not verified accounts, but align with IRA users.

%%Partial conclusion on the caractherization
Among the various groups of accounts, our analyses
reveal that IRA accounts and suspended accounts share similar interests
in terms of the news outlets they reference, with suspended accounts showing a
higher interest in fake and extreme bias right related content. 
Both groups also demonstrate a similar use of non-official clients,
though with differences in the information transmitted through them.
Not very dissimilar are not found accounts, which, however, 
display very low usage of non official clients.

Notably, our analyses uncover parallels in the behavior
of IRA and suspended accounts concerning out-degree activity
in each category. These resemblances might signify a shared
effort by suspended accounts to steer Twitter discourse toward
the right political agenda. However, it is crucial to emphasize
that while this similarity with IRA does not imply
collaboration, it is highly improbable that the entire set
of suspended accounts is involved in this endeavor. Therefore,
identifying a representative subset of suspended accounts
participating in this intent becomes imperative.

\subsection{IRA ego network}
\label{ego network}

To pinpoint a representative subset of suspended
accounts employing strategies akin to IRA accounts, 
we look into the IRA ego network. This network encompasses all users
interacting with IRA through retweets, mentions, quotes, or replies
(see Section \ref{retweet}). Table~\ref{table:four} summarizes the key
features of these networks.

We analyze the interactions of users with different account types,
such as verified, not verified, suspended, and not found. Specifically,
we focus on the top  1000 users involved in each interaction type 
(retweet, reply, quote, and tweet) by considering both in and
out degrees. For instance, in the case of retweet interaction, we consider the top 1000
(most retweeted) non-IRA users who were retweeted by IRA accounts in
the ``out" direction. Similarly, in the ``in" direction, we examine the
top 1000 (most retweeting) non-IRA users who retweeted IRA accounts.

\begin{figure}[!ht]
  \centering
  \subfloat{{\bf (a)}\includegraphics[width=0.40\linewidth]{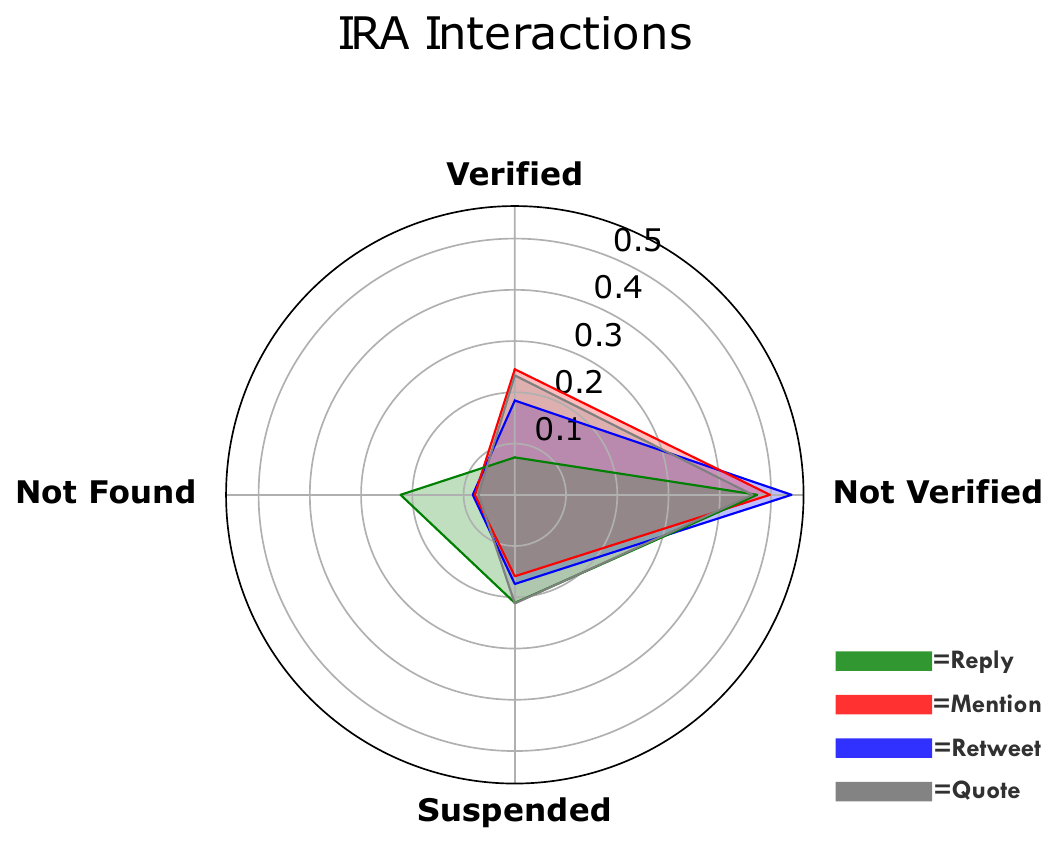}}
  \subfloat{{\bf (b)}\includegraphics[width=0.40\linewidth]{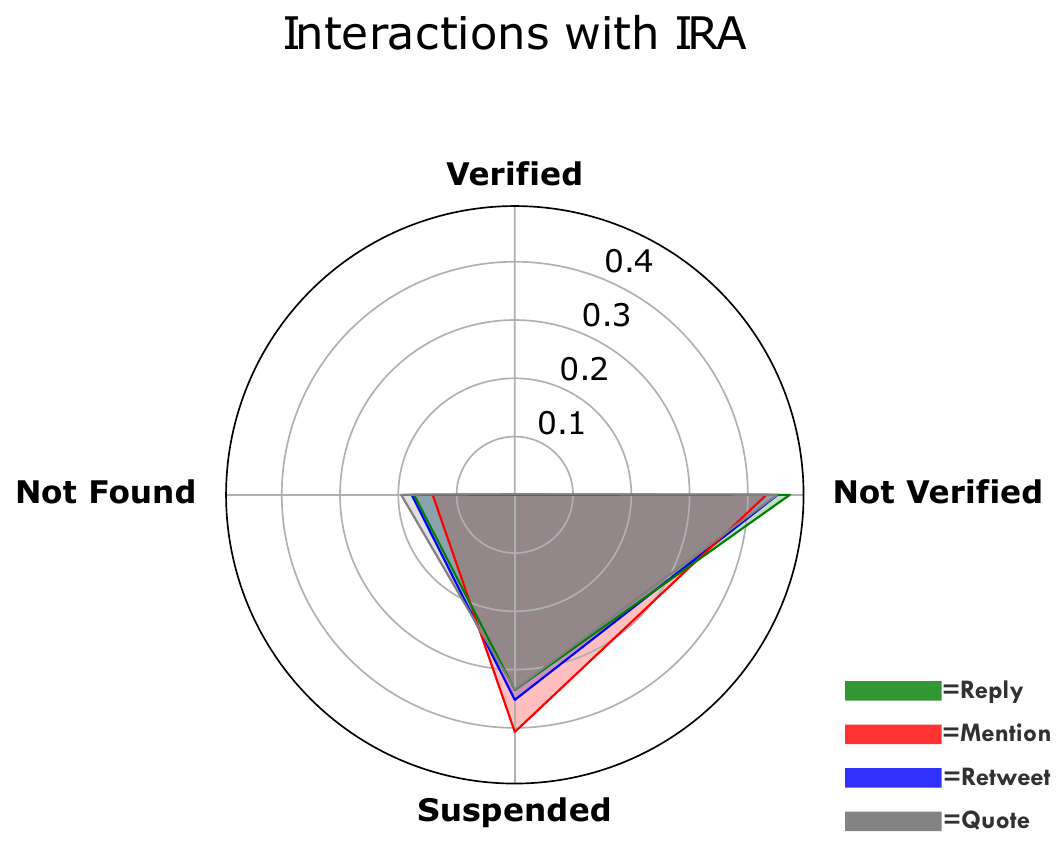}}
  \caption{{\bf Account status: top 1000 active users.} We show the
  fraction of the top 1000 active users per status (in \%) in each interaction network 
  (retweet in blue, mention in red, reply in green, and quote in gray). Possible statuses
  are: Verified, Suspended,
  Not Verified and Not Found. {\bf(a)} ``out'' direction: users are retweeted, 
  mentioned, quoted, or replied to by IRA users. On average, 18.4\% of these users
  have verified accounts, 49.4\% are not verified, 18.9\% are suspended, and 11.4\%
  are not found accounts {\bf(b)} ``in'' direction: users retweet, mention, 
  quote, or reply to IRA users. On average, 45.1\% of them are not verified,
  35.7\% is suspended, and 17.1\% is not found.}
  \label{fig:6}
\end{figure}

Figure~\ref{fig:6}a displays the distribution of the
users interacted by IRA (connections go from the non-IRA users to IRA) 
into the different account types. On average, 18.4\% of these users have 
verified accounts, 49.4\% are not verified, 18.9\% are suspended accounts, and 
11.4\% are not found accounts. Notably, among the verified accounts, we identified the
official profile of President Donald Trump and popular news outlets such as 
The Guardian and FOX NEWS. See Supplementary Tables 9, 10, 11, and 12 for a list of
the top 20 accounts. Figure~\ref{fig:6}b, displays the distribution of the
users who interact with IRA (connections go from the IRA to non-IRA users).
None of the top 1000 users who engage with IRA have verified accounts. 
On average, 45.1\% of them are not verified, 35.7\% is suspended,
and 17.1\% is not found. See Supplementary Tables 13,14, 
15 and 16 for a list of the top 20 accounts.

It is worth noting that the number of suspended accounts interacting with IRA
(amounting to 30,622) is nearly 60 times larger than the number of IRA accounts. 

It is noteworthy to observe that the majority of active users
interacting with IRA are suspended users. Moreover, given that
the most common type of interaction is retweeting, as indicated
in Table~\ref{table:four}, this suggests that most users tend
to retweet IRA, likely with the intent to disseminate similar
types of information. This ultimately confirms the notion that
suspended accounts and IRA share similar views.

{\bf Ego polarization} Analyzing user preferences in terms 
of political orientations (see Sections \ref{supporters} 
and \ref{classes}) reveals, as expected,
a higher presence of right supporters in the IRA ego network.
Specifically, 8.1\% of users are strong supporters of Trump, 
4\% are strong supporters of Clinton, 64.5\% are weak supporters
of Trump, 21.1\% are weak supporters of Clinton, and the
remaining 2.3\% of users are categorized as undecided. {\color{blue}{It's important to note that when calculating user polarization in computing, we consider all users, regardless of their account status. The majority of suspended accounts are classified as Trump supporters, either weak or strong. This is detailed in Table \ref{table:five}, where we also provide the percentages broken down by account status.}}

This aligns with the notion that this subset of suspended accounts was
oriented toward the right agenda. Additional details about the users'
classifications can be found in Supplementary Section 4.

It is also worth noticing that we found a substantial number of IRA accounts 
classified as weak Clinton supporters. This suggests a dual strategy 
employed by the IRA in their campaign, as already suggested in the existing
literature \cite{stewart2018examining,linvill2020troll}.
One aspect of this strategy involves reinforcing the opinions
of users classified as strong Trump supporters. On the other hand,
another set of IRA accounts aims to expand their reach within
left-leaning accounts by mentioning verified accounts classified as
weak and strong Trump supporters.

{\bf Community structure} The exploration of this
dual strategy and the role played by suspended accounts
in it continues through the identification of communities
within the aggregated IRA ego network. This network is
formed by merging the networks for the four types of interaction,
as detailed in Section \ref{retweet}.

We perform multiscale community detection to the largest connected component
of the undirected weighted version of the aggregated network, which contains
99.9\% of the initial nodes. To assess community stability, we utilize Markov
stability, as detailed in Supplementary Section 5. 

We reveal the existence of two prominent communities,
a finding that adds intrigue as it aligns with the notion
that the IRA orchestrated two distinct micro-campaigns. These communities reflect 
the polarization of users. In the biggest community, 76.6\% of users
are classified as weak Trump supporters and 9.5\% as strong Trump supporters,
while the second biggest community left has 70.2\% of weak
Clinton supporters users and 13\% of strong Clinton supporters users, 
see Table~\ref{table:eight}. These results align with the existing
literature \cite{stewart2018examining,linvill2020troll}. Moving
forward, we will refer to these communities as the right and left communities.

The distinction in the political orientation of these two communities
is further supported by analyses of the hashtag clouds constructed
from the content shared by users, as discussed in Supplementary
Section 6. 
Users in the right community tend to share hashtags in support of Trump and
against Clinton, while the opposite holds true for the left community.

Within the right community, suspended accounts comprise 20\% of the total
nodes in the community. This percentage decreases by more than half in
the left community (refer to Table~\ref{table:eight}).
This outcome suggests a difference in strategies
between suspended accounts and IRA, with IRA implementing a dual
strategy (targeting both right and left users), while suspended
accounts predominantly focus on targeting right-oriented users.
These suspended users account for 21.7\% of the interactions, 
with the directions of the connections going from IRA to suspended accounts,
as indicated in Supplementary Figure 7 and Supplementary Table 20. 
In the case of the left community, the type of interaction is inverted, 
and in most cases, it is the IRA engaging with the other groups,
particularly not verified and verified, with the connections
going from suspended to IRA accounting for 12.5\%
of the overall connections in the community.

% Partial conclusion

The above-presented results demonstrate that utilizing
the IRA ego network as a proxy to identify a contained
group of suspended accounts aligning with the right ideologies
is an effective strategy. We find that the majority of these
suspended users are classified as Trump supporters. By examining
the dual strategies employed by the IRA campaign, we reveal that
the majority of suspended accounts were following the strategies
of the right community, specifically targeting Trump supporters.
The directions of the connections suggest that this was primarily
done through interactions with IRA accounts, including retweets,
mentions, quotes, and replies.

{\bf Expanded ego network}
The IRA ego network serves as a proxy for identifying suspended accounts that
share similarities with IRA accounts. However, this does not ensure that
this subset covers all suspended accounts involved in promoting right-related
content. Additional suspended accounts may exist in the
dataset, but rather than interacting directly with IRA, they might be
interacting with the suspended accounts that are engaging with IRA.
To investigate this possibility, we expanded the IRA aggregated
network from the earlier section by incorporating interactions
involving this subset of suspended nodes and all other users.
In essence, we created the ``suspended+IRA" ego network. This network,
akin to the previous section, was constructed based on the four types
of Twitter interactions. Supplementary Table 17 provides comprehensive
information regarding each interaction network.

The inclusion of both IRA accounts and suspended accounts 
significantly amplifies the dimension of the aggregated network,
increasing the number of nodes from 179,783 in the IRA ego network to 1,723,477
in the expanded ego network. This expanded network exhibits 45 times more
connections than the aggregated ego network, with an average degree of $\langle k \rangle = 11$.
Similar to the IRA ego network, retweeting and mentioning interactions
remain the most common types of interactions in the expanded network.

Next, we performed a multi-scale community detection analysis and explored different 
parameter values to identify the optimal partition.  The resulting partition
(resulting in two communities) preserves the communities' polarization, as shown in
Table~\ref{table:nine}, with the two expanded communities
being mostly composed of supporters of Trump and Clinton. When scrutinizing
the composition of these two communities,
it becomes evident that they are predominantly comprised of not verified,
verified, and not found accounts. Suspended accounts, despite
constituting a smaller percentage compared to the IRA ego network,
also exhibit significantly lower activity. Their number of
connections in both directions represent less than 5\% of
the total connections, as illustrated in Table~\ref{table:nine}.

These findings indicate minimal and negligible
interactions between suspended accounts within the IRA ego network
and other suspended accounts. This emphasizes that the suspended
accounts identified in the IRA ego network represent the most
significant group of suspended accounts disseminating right-related information,
similar to IRA, in the Twitter discourse.

\subsection{Causal network patterns: IRA nodes versus suspended nodes}
\label{Section4}

This section investigates the impact that suspended accounts and IRA 
have on shaping the Twitter discourse during the 2016 US presidential elections. 
{\color{blue}Specifically, we scrutinize the causal relationships between IRA
(and suspended accounts) tweet activity and the activity of supporting classes,
namely weak Trump supporters, weak Clinton supporters, strong Trump supporters,
strong Clinton supporters, and undecided users.}

We employ a multivariate Granger causal network reconstruction approach
to establish links between the activity of IRA (suspended) nodes
and the supporting classes. This is achieved using the causal discovery 
algorithm \cite{spirtes2000causation,runge2012escaping,runge2019detecting}, 
which tests the independence of each pair of time series for several time 
lags conditioned on potential causal parents using a Partial Correlation
Independence test and it removes spurious correlations. We use the algorithm for
variable selection and perform a linear regression using 
only the true causal link discovered. We choose linear causal effects for their 
reliability and interpretability, which allows us to compare causal effects as 
first-order approximations, estimate the uncertainties of the model, and 
construct a causal-directed weighted network \cite{runge2015identifying}.
The causal effect between a time series $X^i$ and $X^j$ at a time delay $\tau$,
$I^{CE}_{i \rightarrow j}(\tau)$, is determined by the expected value of $X^j_t$ 
(in units of standard deviation) if $ x^i (t - \tau)$ is perturbed by one standard 
deviation \cite{runge2015identifying,bovet2019influence}. However, an assumption
of causal discovery is causal sufficiency,
which assumes that every common cause of any two or more variables
is present in the system \cite{spirtes2000causation}. In our case,
causal sufficiency is not satisfied because Twitter's activity is
only a part of a larger social system. Therefore, the term ``causal"
should be understood as relative to the system under study \cite{bovet2019influence}.

We created time series of Twitter activities by counting the number of
tweets posted by each node belonging to one of the supporting classes at a
15-minute resolution. We only consider users that belong to the verified and 
not verified classes, and only consider the tweets coming from official clients.
Instead, for the IRA (suspended) nodes, we consider all the tweets, no matter the clients.
To remove trend and circadian cycles from the time series, we utilized
the STL (seasonal trend decomposition procedure based on Loess) method \cite{cleveland1990stl},
which decomposes a time series into seasonal (in this case, daily),
trend, and remainder components. We used the residuals of the STL
filtering of the 15-minute tweet volume time series. 

In simpler terms, Granger causality examines
whether past retweet behaviors in one group can assist in
forecasting the retweet behaviors of the other group.
It doesn't imply a direct cause-and-effect relationship but rather
investigates whether changes in one group's activity precede
changes in the other group's activity.

Tables~\ref{table:eleven} and \ref{table:twelve} present the causal relationships
among different groups in the two scenarios: one with only IRA nodes and the other
with suspended nodes. The direction of each link is from the column group to the row group.
For example, considering the strong Trump supporters, their causal effect on the weak
Clinton supporters is measured at 0.16 $\pm$ 0.011, as shown in Table~\ref{table:eleven}.
The blue entries in the tables represent the auto-correlation of each time series.
In both scenarios, the auto-correlations exhibit the strongest causal effects for all
time series, except for the undecided group.

\begin{figure}[ht!]
  \centering
  \subfloat{{\bf (a)} \includegraphics[width=0.45\linewidth]{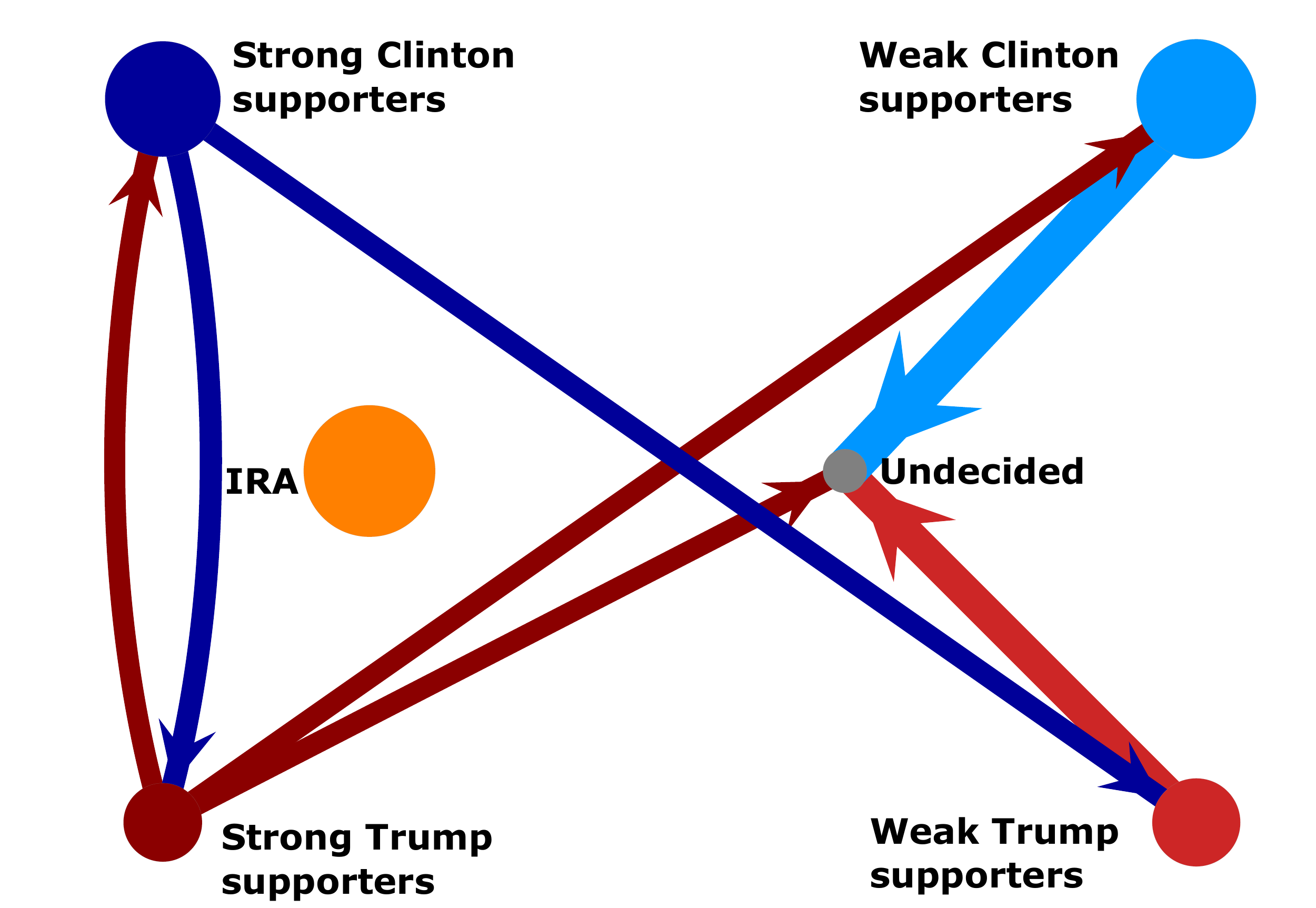}}
  \subfloat{{\bf (b)} \includegraphics[width=0.45\linewidth]{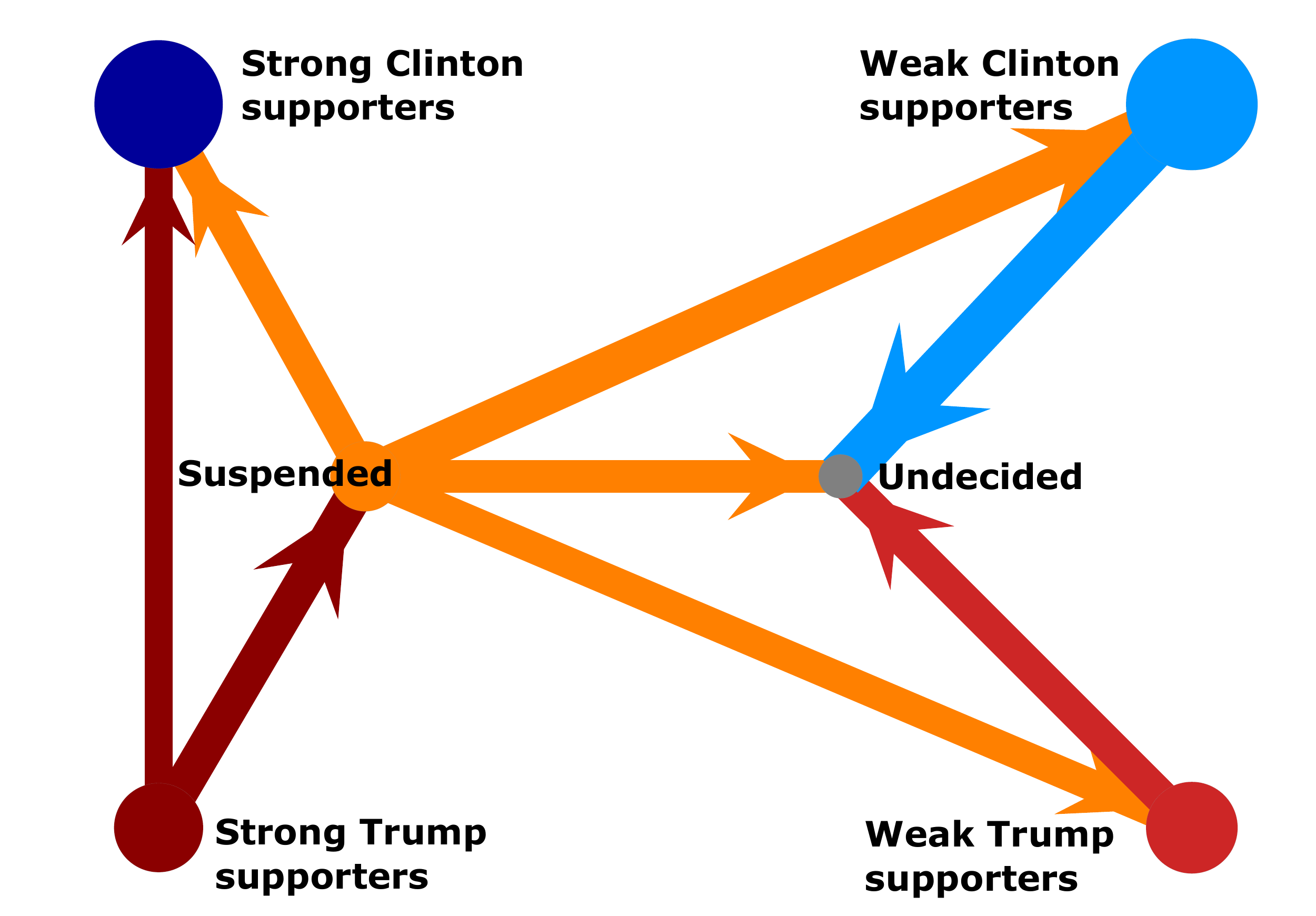}}
  \caption{{\bf Causal Networks.} {\bf (a)}  Graph showing the maximal causal effects between the activity of the IRA nodes and the supporting classes of the presidential candidates. Arrows indicate the direction of the maximal causal effect ($\geq 0.16$) between two activity time series. The width of each arrow is proportional to the strength of the causation, and the size of each node is proportional to the auto-correlation of each time series. Dark blue and dark red highlight the contribution of strong Clinton and Trump supporters, respectively. Light blue and light red are associated with the weak Clinton and Trump supporters, gray with the undecided users, and orange with the IRA nodes. The causal relation primarily flows from strong supporters of both Trump and Clinton to weak and strong supporters of opposing political candidates. Additionally, weak supporters from both sides play a role in influencing the undecided group, with weak Trump supporters receiving support from strong Trump supporters in their efforts. Notably, IRA nodes do not play a significant role in this causal network, suggesting that they have limited influence
on shaping Twitter discourse. {\bf (b)} Graph showing the maximal causal effects between the activity of the suspended nodes and the supporting classes of the presidential candidates. Arrows indicate the direction of the maximal causal effect ($\geq 0.2$) between two activity time series.  Strong Trump supporters have a causal effect on Suspended nodes, which,
in turn, have a causal influence on both weak supporters and the undecided group.
Additionally, weak supporters continue to exert a causal effect on the undecided group. Strong Trump supporters have a causal effect on strong Clinton
supporters, but not vice-versa.} 
  \label{fig:13}
\end{figure}

To identify the most significant causal links, a threshold of 0.16 (0.20 for the
suspended scenario) was set on the causal relation, selecting connections that account
for 75\% of the total effect. These selected links are highlighted in bold in the
tables. Figures~\ref{fig:13}a and \ref{fig:13}b visualize the causal networks 
constructed using these connections. The nodes are colored as follows: dark 
red for strong Trump supporters, dark blue for strong Clinton supporters, light 
red for weak Trump supporters, light blue for weak Clinton supporters, orange 
for the IRA nodes, and gray for the undecided group. Arrows indicate the direction
of maximal causal effect ($\geq$0.16 and $\geq$ 0.20) between two activity time series. The
width of each arrow represents the strength of the causation, and the size of each 
node is proportional to the auto-correlation of each time series.

Figures~\ref{fig:13}a and b, illustrate contrasting scenarios 
in terms of the causal network structure when considering IRA nodes
alone versus suspended nodes. In Figure~\ref{fig:13}a, which represents the causal 
network considering IRA nodes only, the influence primarily flows from strong 
supporters of both Trump and Clinton to weak and strong supporters of
opposing political candidates. Additionally, weak supporters from both sides
play a role in influencing the undecided group, with weak Trump supporters
receiving support from strong Trump supporters in their efforts. Notably, IRA 
nodes do not play a significant role in this causal network, suggesting that 
they have limited causation on users' activity. 

On the other hand, in Fig.~\ref{fig:13}b, which represents the causal network for suspended
nodes, the structure shows substantial differences. Suspended nodes take on a central role,
acting as a bridge between strong Trump supporters and the weak and undecided supporters.
Strong Trump supporters have a causal effect on suspended nodes, which, in turn,
have a causal influence on both weak supporters and the undecided group. Additionally,
weak supporters continue to exert a causal effect on the undecided group. Interestingly,
strong Trump supporters have a causal effect on strong Clinton supporters, but not vice-versa.

\section{Discussions and Conclusions}
\label{Section5}

Current research focuses on the role and impact of the Internet
Research Agency (IRA) in the 2016 US presidential elections.
This emphasis on IRA's political interference may have overshadowed
other campaigns with similar aims that were not linked to Russian origins.
By merging the IRA public dataset with a collection of tweets spanning
the five months leading up to the 2016 presidential elections, our
objective is to investigate the presence and impact of suspended accounts—those
not flagged as IRA—which might have contributed to the dissemination
of content aligned with the Trump political agenda.

Our analysis reveals that the IRA and suspended accounts (not flagged as IRA by Twitter)
do share many similarities, in terms of the type of news they share, the clients they use
and the way participate in the Twitter social discourse, as highlighted in Section \ref{characterization}. However, expecting all the suspended accounts in our extensive dataset, comprising over 700,000 users (7.7\% of the 10 million users, as detailed in Section \ref{data}),
to exhibit such similarity is improbable. To pinpoint a representative group more akin to IRA, we leverage the IRA ego network.

Within the IRA ego network, we identified 30,622 suspended accounts,
a number 60 times larger than the IRA accounts. These
suspended accounts engaged through various interactions
like retweeting, mentioning, replying, and quoting, with retweets
and mentions being the most common. Aligning with existing
literature \cite{golovchenko2020cross, linvill2019russians}
that asserts the IRA aimed to support Donald Trump and sow discord in the U.S.,
we found that the majority of nodes in the ego network,
including suspended accounts, are classified as Trump supporters
(discussed in Section \ref{ego network}).

In the aggregated IRA ego network, approximately 2\% of total users
were directly exposed to IRA content, consistent with \cite{eady2023exposure}.
A multiscale community detection on this network revealed two communities,
encompassing almost 90\% of the total nodes, indicating user polarization.
The larger community (community right) aimed to support Trump, while
the smaller one (community left) interacted directly with Clinton supporters,
potentially attempting to influence their opinions. This
dual strategy aligns with current literature suggesting
a multifaceted approach by the IRA \cite{stewart2018examining, linvill2020troll}.

Contrary to the multifaceted strategy observed with the IRA,
suspended accounts exhibit a more focused role in the right community.
They contribute significantly, comprising over 20\% of the connections
in the community, with a majority of these connections originating from
IRA to suspended accounts. Building on earlier findings, this suggests
that the suspended accounts in this group were primarily engaged in
mentioning and retweeting IRA content, likely aiming to inundate the
social platform with ideas consistent with the right political agenda.

Having identified a group of suspended accounts resembling IRA
behaviors (partially) and gaining insights into their intent,
we proceed to measure the impact of both suspended accounts and IRA
on shaping the Twitter discourse during the 2016 US presidential elections.
{\color{blue}This is achieved through the application of Granger causality to the
tweet activity produced by IRA and suspended accounts, and each supporting
class (refer to Section \ref{Section4}).}

Our causal analyses reveal that the group of IRA accounts did not have a 
significant impact in influencing the candidate's supporters, as shown in Figure
\ref{fig:13}a and detailed in \cite{eady2023exposure}.
However, the situation becomes more intricate when we consider the
suspended accounts. We find that these users wielded a 
substantial influence on individuals categorized as undecided or weak supporters,
potentially with the intention of swaying their opinions. This effect 
is graphically portrayed in Figure \ref{fig:13}b, illustrating the 
bridging effect that suspended nodes played between strong Trump supporters
and the group of weak supporters and undecided individuals. 

It's important to note that while Granger causality suggests
that past retweet behaviors in one group can aid in predicting 
the retweet behaviors of the other group, it doesn't imply a direct
cause-and-effect relationship. The determination of such causality
goes beyond the scope of this study. Additionally, utilizing tweet
activity provides insights into user behavior, but conclusions
regarding changes in users' vote intentions require longitudinal data. {\color{blue}{It is, however, noteworthy that the similarity observed between the Internet Research Agency (IRA) and the group of suspended users, coupled with the fact that suspended accounts influenced the activity of undecided users, opens up the possibility of a new scenario, such as potential cooperation between the IRA and the identified group of suspended users. It's also conceivable that this group of suspended accounts was part of the IRA's campaign and remained undetected by Twitter. However, this remains purely speculative, and further analysis and data are needed to draw more concrete conclusions.}}

Furthermore, the lack of detailed information about the nature of
suspended accounts, such as whether they are trolls or bots, is a
limitation. While all possibilities are considered, the logistical
challenges of controlling a group of over 30,000 accounts make
it more likely that this set of suspended accounts predominantly
consists of bots. 

In summary, this study suggests a scenario in which a
significant group of suspended accounts, often overshadowed by
the IRA narrative, played a crucial role during the 2016 US
presidential elections. Further research is required to better
understand their impact on political user preferences.

\section*{List of abbreviations}
% If abbreviations are used in the text they should be defined in the text at first use, and a list of abbreviations should be provided.
IRA: refer to the Internet Research Agency.

\section*{Declarations}

\subsection*{Availability of data and materials}
The Twitter data are provided according to its terms and are available at \url{https://osf.io/g4hws/} and \url{https://github.com/makselab/IRA-and-suspended-accounts.}.
Analytical codes are available in the same repositories. 

\subsection*{Competing interests}
The authors declare that they have no competing interests.

\subsection*{Funding}
HAM was supported by NSF-HNDS Award 2214217. JSA gratefully acknowledges the Brazilian agencies FUNCAP, CNPq and CAPES, the National Institute of Science and Technology for Complex Systems in Brazil, and the PRONEX-FUNCAP/CNPq Award PR2-0101-00050.01.00/15 for financial support. MS gratefully acknowledges the Brazilian agency CAPES Award 88887.899221/2023-00. ZZ was supported by the National Natural Science Foundation of China under project No. 62302319 and R\&D Program of Beijing Municipal Education Commission (Grant No. KM202210038002).

\subsection*{Authors' contributions}
HAM, AB and MS conceived the research; MS and AB designed and supervised the
research; MS and AB coordinated and supervised the analysis; MS and ZZ performed the analyses.
MS, ZZ, JSA, AB, and HAM analyzed the results; MS and ZZ wrote the first draft. All authors
edited and approved the paper. 

\subsection*{Acknowledgements}
 Not applicable.

\clearpage

\begin{table}[!ht]
\center
\resizebox{\textwidth}{!}{\begin{tabular}{|l|r|r|r|r|r|r|l|}\hline
 & \multicolumn{3}{c|}{Full Network} & \multicolumn{3}{c|}{IRA} \\
 &  N  &  E  & $\langle k \rangle /2$ &  $N_{IRA}$ & $\langle k_{out}\rangle$  & $\langle k_{in} \rangle$ \\\hline
Fake               &       175,605 &  1,143,083 &    6.5 &      54 &    32.5 &    5.4 \\
Extreme bias right &       249,659 &  1,637,927 &    6.6 &      70 &    27.3 &    7.1 \\
Right              &       345,644 &  1,797,023 &    5.2 &      84 &    30.1 &    4.7 \\
Right leaning      &       216,026 &   495,307 &    2.3 &      67 &     6.3 &    1.7 \\
Center             &       864,733 &  2,501,037 &    2.9 &     163 &     4.6 &    2.6 \\
Left leaning       &      1,043,436 &  3,570,653 &    3.4 &     140 &     8.9 &    2.2 \\
Left               &       536,903 &  1,801,658 &    3.4 &     105 &     4.3 &    2.4 \\
Extreme bias Left  &        78,911 &   277,483 &    3.5 &      10 &     0.0 &    1.6 \\\hline
\end{tabular}}
\caption{{\bf Retweet categories' networks.} The table contains the characteristics of each of the eight 
retweet networks, such as the number of nodes $N$, the number of edges $E$, and the average degree
$\langle k_{in} \rangle = \langle k_{out} \rangle = \langle k\rangle/2 $. We also report the number
of IRA users in each retweet network $N_{IRA}$, as well as their average in-degree $ \langle k_{in} 
\rangle$ and out-degree $\langle k_{out} \rangle$.}
\label{table:one}
\end{table}

\begin{table}[!ht]
\center
\resizebox{\textwidth}{!}{\begin{tabular}{|l|c|c|c|c|c|l|}\hline
& \multicolumn{3}{c|}{Full networks} & \multicolumn{3}{c|}{IRA users}\\
           & N & E & $ \langle k \rangle$   & $N_{IRA}$ & $ \langle k_{out} \rangle$ & $\langle k_{in} \rangle$\\\hline
Retweeting &  154,366  &   360,265  &  2.3 &   497  &   468.4  & 262.0 \\
Mentioning &  70,926   &   197,644  &  2.8 &   508  &   353.1  & 41.7 \\
Replying   &  14,225   &   16,775   &  1.2 &   193  &   71.2   & 16.4 \\
Quoting    &  19,195   &   31,538   &  1.6 &   353  &   36.4   & 53.6 \\ \hline
Aggregated &  179,783  &   432,429  &  2.4 &   524  &   486.8  & 343.8\\\hline 
\end{tabular}}
\caption{{\bf Interactions ego networks.} The table contains information about each type of interaction network, as well as 
information about their aggregated version. We report the number of nodes $N$, edges $E$, the average degree $<k>$, 
and the number of IRA nodes $N_{IRA}$, with their in/out-degree. Retweeting and mentioning are the two most frequent types
of interactions between IRA and non-IRA users.  }
\label{table:four}
\end{table}

\begin{table}[!ht]
\center
\resizebox{\textwidth}{!}{
\begin{tabular}{|l|r|r|r|r|r|l|}\hline
 &    Strong Trump supporters  &  Weak Trump supporters &  Strong Clinton supporters & Weak Clinton supporters  &   Undecided \\\hline
Not Found    &   8.13 &  73.44 &   2.17 &  14.49 &  1.76 \\
Not Verified &   6.66 &  63.79 &   4.20 &  22.94 &  2.41 \\
Suspended    &  14.43 &  67.93 &   3.56 &  11.78 &  2.30 \\
Verified     &   0.25 &  15.84 &  11.63 &  68.95 &  3.33 \\
IRA          &   0.32 &  34.62 &   0.00 &  60.26 &  4.81 \\\hline
\end{tabular}}
\caption{{\bf Groups classes.} The table presents the distribution
of different user types, namely not found, not verified, suspended, verified, and IRA users,
among the various supporting classes. The percentages in the table are normalized per account type,
meaning that the sum of percentages of a given account type for each supporting class adds up to 100\%.}
\label{table:five}
\end{table}

\begin{table}[!ht]
\centering
\begin{tabular}{|l|r|r|l|}\hline
  &  Community right & Community left \\\hline
n° nodes                 &  135,846 &   24,318 \\
n° IRA nodes             &     160 &     114 \\\hline
Undecided             &       1.8 \% &       4.1 \% \\
Strong Trump supporters            &       9.5 \% &       2.1 \% \\
Weak Trump supporters              &      76.6 \% &      10.5 \% \\
Strong Clinton supporters            &       1.8 \% &      13.0 \% \\
Weak Clinton supporters             &      10.3 \% &      70.2 \% \\\hline
Not Found      &      21.2 \%  &      11.6 \%  \\
Not Verified   &      56.6 \%  &      63.8 \%  \\
Suspended      &      20.3 \%  &       9.6 \%  \\
Verified       &       1.7 \%  &      14.5 \%  \\
IRA            &       0.1 \%  &       0.5 \%  \\\hline
\end{tabular}
\caption{{\bf IRA ego network: partition characteristics.} Characteristics of the 
communities in the IRA ego network. We display information for the
communities with at least 10\% of the nodes of the overall
network. For each community, we report the number of nodes, the number of IRA accounts,
the share of supporting classes and the distribution of users among different groups.}
\label{table:eight}
\end{table}

\begin{table}[!ht]
\centering
\begin{tabular}{|l|r|r|l|}\hline
  &  Community E-right & Community E-left \\\hline
n° nodes                 &  718,825 &   314,559 \\
n° IRA nodes             &    25,565 &     4,359 \\\hline
Undecided              &       5.4 \% &        4.1 \% \\
Strong Trump supporters            &      10.2 \% &        2.6 \% \\
Weak Trump supporters            &      55.9 \% &       16.0 \% \\
Strong Clinton supporters             &      3.7 \% &       10.6 \% \\
Weak Clinton supporters              &       24.8 \% &       66.6 \% \\\hline
Not Found      &      19.8 \% &       15.9 \% \\
Not Verified   &       60.2 \% &       69.8 \% \\
Suspended      &     14.2 \% &        8.9 \% \\
Verified       &      2.2 \% &        4.0 \% \\
IRA            &        3.6  \% &        1.4 \% \\\hline
\end{tabular}
\caption{{\bf Expanded ego network: partition characteristics.} Characteristics of the 
communities. We display information for the communities with at least 10\% of the nodes of the overall
network. For each community, we report the number of nodes, the number of IRA accounts,
the share of supporting classes, and the distribution of users among different groups.}
\label{table:nine}
\end{table}

\begin{table}[!ht]
\resizebox{\textwidth}{!}{ 
\begin{tabular}{|l|r|r|r|r|r|r|l|}\hline
      &    IRA &       Weak Trump supporters &        Weak Clinton supporters  &    Strong Trump supporters   &    Strong Clinton supporters  &   Undecided    \\\hline
IRA   &   {\textcolor{blue}{ 0.87  $\pm$ 0.006}} &  0  &  0  &  0  & 0  & 0   \\
Weak Trump supporters    &   0  &   {\textcolor{blue}{0.54  $\pm$  0.0155}} &  0.08  $\pm$ 0.024 &  0.15  $\pm$ 0.015  &  {\bf 0.16  $\pm$ 0.022} &  0.05 $\pm$  0.019  \\
Weak Clinton supporters     &   0  &  0.1  $\pm$  0.021 &   {\textcolor{blue}{0.78  $\pm$ 0.014}} &  {\bf 0.16  $\pm$ 0.011}  &  0.13  $\pm$ 0.02  &  0.09  $\pm$  0.02   \\
Strong Trump supporters    &   0  &  0.01  $\pm$  0.004 &  0.03  $\pm$ 0.015 &   {\textcolor{blue}{0.47  $\pm$  0.011}}  &   {\bf  0.17  $\pm$ 0.015}  &  0.01  $\pm$  0.003  \\
Strong Clinton supporters   &   0  &   0.06  $\pm$ 0.018 &  0.09  $\pm$ 0.013 &  {\bf 0.16  $\pm$  0.012}  &   {\textcolor{blue}{0.75  $\pm$ 0.013}}  &  0.07  $\pm$ 0.016  \\
Undecided     &   0  &   {\bf 0.26  $\pm$ 0.017} &  {\bf0.34  $\pm$ 0.021} &  {\bf 0.16  $\pm$ 0.015}  &  0.14  $\pm$  0.02  &   {\textcolor{blue}{0.2  $\pm$ 0.019}} \\\hline
\end{tabular}}
\caption{{\bf Causal Links: IRA.} We show the value of the maximal causal effect,
$I_{i \to j}^{{\mathrm{CE, max}}} = {\mathrm{max}}_{0 < \tau \le \tau _{{\mathrm{max}}}}\left| {I_{i \to j}^{{\mathrm{CE}}}(\tau )} \right|$ between each pair (i, j) of activity time series, 
 where $\tau_{max}= 18 \times 15$min=4.5h is the maximal time 
lag considered, with standard errors. The arrows indicate the 
the direction of the causal effect. For each activity time series,
we indicate in bold the most important drivers of activity (excluding themselves).
In blue, we highlight the auto-correlation of each node. }
\label{table:eleven}
\end{table}

\begin{table}[!ht]
\resizebox{\textwidth}{!}{ \begin{tabular}{|l|r|r|r|r|r|r|l|}\hline
      &   Suspended &   Weak Trump supporters &   Weak Clinton supporters &   Strong Trump supporters   &    Strong Clinton supporters  &   Undecided \\\hline
Suspended  &  {\textcolor{blue}{0.38 $\pm$ 0.012}}  &  0.05 $\pm$ 0.008  &  0.04 $\pm$ 0.012  &  {\bf 0.27 $\pm$ 0.018}  &  0.17 $\pm$ 0.020   &  0.03 $\pm$ 0.007  \\
Weak Trump supporters    &  {\bf 0.23 $\pm$ 0.027}  & { \textcolor{blue}{ 0.52 $\pm$ 0.015 }}  &  0.08 $\pm$ 0.021  &  0.10 $\pm$ 0.027  &  0.19 $\pm$ 0.023   &  0.05 $\pm$ 0.019  \\
Weak Clinton supporters    &   {\bf 0.29 $\pm$ 0.035}  &  0.06 $\pm$ 0.017  & { \textcolor{blue}{0.79 $\pm$ 0.014 }}  &  0.19 $\pm$ 0.037  &  0.12 $\pm$ 0.021   &  0.07 $\pm$ 0.019  \\
Strong Trump supporters      &   0.05 $\pm$ 0.009  &  0.03 $\pm$ 0.011  &  0.02 $\pm$ 0.012  & { \textcolor{blue}{0.50 $\pm$ 0.011 }}  &  0.15 $\pm$ 0.017   &  0.05 $\pm$ 0.012  \\
Strong Clinton supporters      &   {\bf 0.23 $\pm$ 0.030}  &  0.02 $\pm$ 0.010  &  0.04 $\pm$ 0.012  &  {\bf 0.20 $\pm$ 0.032}  & {\textcolor{blue}{ 0.77 $\pm$ 0.012 }}   &  0.04 $\pm$ 0.006  \\
Undecided    &  {\bf 0.24 $\pm$ 0.027}  & {\bf 0.25 $\pm$ 0.015 } &  {\bf 0.34 $\pm$ 0.021}  &  0.10 $\pm$ 0.027  &  0.16 $\pm$ 0.020   &  {\textcolor{blue}{ 0.20 $\pm$ 0.019 }}   \\\hline
\end{tabular}}
\caption{{\bf Causal Links: Suspended.} We show the value of the maximal causal effect,
$I_{i \to j}^{{\mathrm{CE, max}}} = {\mathrm{max}}_{0 < \tau \le \tau _{{\mathrm{max}}}}\left| {I_{i \to j}^{{\mathrm{CE}}}(\tau )} \right|$ between each pair (i, j) of activity time series, 
 where $\tau_{max}= 18 \times 15$min=4.5h is the maximal time 
lag considered, with standard errors. The arrows indicate the 
the direction of the causal effect. For each activity time series,
we indicate in bold the three most important drivers of activity (excluding themselves).
In blue, we highlight the auto-correlation of each node. }
\label{table:twelve}
\end{table}

\clearpage
\newpage
%\bibliographystyle{naturemag}
%\bibliography{bibliography}
%\bibliographystyle{elsarticle-num}
%\bibliography{bib-elite.bib}
\bibliography{sn-bibliography}% common bib file

\begin{thebibliography}{10}
\expandafter\ifx\csname url\endcsname\relax
  \def\url#1{\burl{#1}}\fi
\expandafter\ifx\csname urlprefix\endcsname\relax\def\urlprefix{URL }\fi
\providecommand{\bibinfo}[2]{#2}
\providecommand{\eprint}[2][]{\url{#2}}
\providecommand{\doi}[1]{\url{https://doi.org/#1}}
\bibcommenthead

\bibitem{digrazia2013more}
\bibinfo{author}{DiGrazia, J.}, \bibinfo{author}{McKelvey, K.}, \bibinfo{author}{Bollen, J.} \& \bibinfo{author}{Rojas, F.}
\newblock \bibinfo{title}{More tweets, more votes: Social media as a quantitative indicator of political behavior}.
\newblock \emph{\bibinfo{journal}{PloS one}} \textbf{\bibinfo{volume}{8}}, \bibinfo{pages}{e79449} (\bibinfo{year}{2013}).

\bibitem{anstead2015social}
\bibinfo{author}{Anstead, N.} \& \bibinfo{author}{O'Loughlin, B.}
\newblock \bibinfo{title}{Social media analysis and public opinion: The 2010 uk general election}.
\newblock \emph{\bibinfo{journal}{Journal of computer-mediated communication}} \textbf{\bibinfo{volume}{20}}, \bibinfo{pages}{204--220} (\bibinfo{year}{2015}).

\bibitem{bovet2018validation}
\bibinfo{author}{Bovet, A.}, \bibinfo{author}{Morone, F.} \& \bibinfo{author}{Makse, H.~A.}
\newblock \bibinfo{title}{Validation of twitter opinion trends with national polling aggregates: Hillary clinton vs donald trump}.
\newblock \emph{\bibinfo{journal}{Scientific reports}} \textbf{\bibinfo{volume}{8}}, \bibinfo{pages}{8673} (\bibinfo{year}{2018}).

\bibitem{ahmed20162014}
\bibinfo{author}{Ahmed, S.}, \bibinfo{author}{Jaidka, K.} \& \bibinfo{author}{Cho, J.}
\newblock \bibinfo{title}{The 2014 indian elections on twitter: A comparison of campaign strategies of political parties}.
\newblock \emph{\bibinfo{journal}{Telematics and Informatics}} \textbf{\bibinfo{volume}{33}}, \bibinfo{pages}{1071--1087} (\bibinfo{year}{2016}).

\bibitem{majo2021role}
\bibinfo{author}{Maj{\'o}-V{\'a}zquez, S.}, \bibinfo{author}{Congosto, M.}, \bibinfo{author}{Nicholls, T.} \& \bibinfo{author}{Nielsen, R.~K.}
\newblock \bibinfo{title}{The role of suspended accounts in political discussion on social media: Analysis of the 2017 french, uk and german elections}.
\newblock \emph{\bibinfo{journal}{Social Media+ Society}} \textbf{\bibinfo{volume}{7}}, \bibinfo{pages}{20563051211027202} (\bibinfo{year}{2021}).

\bibitem{hegelich2016social}
\bibinfo{author}{Hegelich, S.} \& \bibinfo{author}{Janetzko, D.}
\newblock \bibinfo{title}{Are social bots on twitter political actors? empirical evidence from a ukrainian social botnet}.
\newblock \emph{\bibinfo{journal}{Proceedings of the International AAAI Conference on Web and Social Media}} \textbf{\bibinfo{volume}{10}}, \bibinfo{pages}{579--582} (\bibinfo{year}{2016}).

\bibitem{ratkiewicz2011detecting}
\bibinfo{author}{Ratkiewicz, J.} \emph{et~al.}
\newblock \bibinfo{title}{Detecting and tracking political abuse in social media}.
\newblock \emph{\bibinfo{journal}{Proceedings of the International AAAI Conference on Web and social media}} \textbf{\bibinfo{volume}{5,1}}, \bibinfo{pages}{297--304} (\bibinfo{year}{2011}).

\bibitem{bruno2022brexit}
\bibinfo{author}{Bruno, M.}, \bibinfo{author}{Lambiotte, R.} \& \bibinfo{author}{Saracco, F.}
\newblock \bibinfo{title}{Brexit and bots: characterizing the behaviour of automated accounts on twitter during the uk election}.
\newblock \emph{\bibinfo{journal}{EPJ Data Science}} \textbf{\bibinfo{volume}{11}}, \bibinfo{pages}{17} (\bibinfo{year}{2022}).

\bibitem{burki2020online}
\bibinfo{author}{Burki, T.}
\newblock \bibinfo{title}{The online anti-vaccine movement in the age of covid-19}.
\newblock \emph{\bibinfo{journal}{The Lancet Digital Health}} \textbf{\bibinfo{volume}{2}}, \bibinfo{pages}{e504--e505} (\bibinfo{year}{2020}).

\bibitem{tucker2017liberation}
\bibinfo{author}{Tucker, J.~A.}, \bibinfo{author}{Theocharis, Y.}, \bibinfo{author}{Roberts, M.~E.} \& \bibinfo{author}{Barber{\'a}, P.}
\newblock \bibinfo{title}{From liberation to turmoil: Social media and democracy}.
\newblock \emph{\bibinfo{journal}{Journal of democracy}} \textbf{\bibinfo{volume}{28}}, \bibinfo{pages}{46--59} (\bibinfo{year}{2017}).

\bibitem{ferrara2017disinformation}
\bibinfo{author}{Ferrara, E.}
\newblock \bibinfo{title}{Disinformation and social bot operations in the run up to the 2017 french presidential election}.
\newblock \emph{\bibinfo{journal}{arXiv preprint arXiv:1707.00086}}  (\bibinfo{year}{2017}).

\bibitem{TheDisinformationReport}
\bibinfo{author}{DiResta, R.} \emph{et~al.}
\newblock \bibinfo{title}{The tactics \& tropes of the internet research agency}.
\newblock \bibinfo{howpublished}{\url{https://www.documentcloud.org/documents/5632786-NewKnowledge-Disinformation-Report-Whitepaper}}.
\newblock \bibinfo{note}{Accessed: 2023-05-25}.

\bibitem{jamieson2020cyberwar}
\bibinfo{author}{Jamieson, K.~H.}
\newblock \emph{\bibinfo{title}{Cyberwar: How Russian Hackers and Trolls Helped Elect a President: What We Don't, Can't, and Do Know}}  (\bibinfo{publisher}{Oxford University Press}, \bibinfo{address}{Oxford}, \bibinfo{year}{2020}).

\bibitem{mueller2019report}
\bibinfo{author}{Mueller, R.~S.} \& \bibinfo{author}{Cat, M. W.~A.}
\newblock \emph{\bibinfo{title}{Report on the Investigation Into Russian Interference in the 2016 Presidential Election}} Vol.~\bibinfo{volume}{1} (\bibinfo{publisher}{US Department of Justice}, \bibinfo{address}{Washington, DC}, \bibinfo{year}{2019}).

\bibitem{carroll2017st}
\bibinfo{author}{Carroll, O.}
\newblock \bibinfo{title}{St. petersburg troll farm had 90 dedicated staff working to influence us election campaign}.
\newblock \emph{\bibinfo{journal}{The Independent}}  (\bibinfo{year}{2017}).

\bibitem{badawy2018analyzing}
\bibinfo{author}{Badawy, A.}, \bibinfo{author}{Ferrara, E.} \& \bibinfo{author}{Lerman, K.}
\newblock \bibinfo{title}{Analyzing the digital traces of political manipulation: The 2016 russian interference twitter campaign}.
\newblock \emph{\bibinfo{journal}{2018 IEEE/ACM international conference on advances in social networks analysis and mining (ASONAM)}} \bibinfo{pages}{258--265} (\bibinfo{year}{2018}).

\bibitem{howard2018ira}
\bibinfo{author}{Howard, P.~N.}, \bibinfo{author}{Ganesh, B.}, \bibinfo{author}{Liotsiou, D.}, \bibinfo{author}{Kelly, J.} \& \bibinfo{author}{Fran{\c{c}}ois, C.}
\newblock \emph{\bibinfo{title}{The IRA, social media and political polarization in the United States, 2012-2018}}  (\bibinfo{publisher}{University of Oxford}, \bibinfo{address}{Oxford}, \bibinfo{year}{2018}).

\bibitem{stewart2018examining}
\bibinfo{author}{Stewart, L.~G.}, \bibinfo{author}{Arif, A.} \& \bibinfo{author}{Starbird, K.}
\newblock \bibinfo{title}{Examining trolls and polarization with a retweet network}.
\newblock \emph{\bibinfo{journal}{Proc. ACM WSDM, workshop on misinformation and misbehavior mining on the web}} \textbf{\bibinfo{volume}{70}} (\bibinfo{year}{2018}).

\bibitem{diresta2018tactics}
\bibinfo{author}{DiResta, R.} \emph{et~al.}
\newblock \bibinfo{title}{The tactics \& tropes of the internet research agency}.
\newblock \emph{\bibinfo{journal}{New Knowledge}}  (\bibinfo{year}{2018}).

\bibitem{zannettou2019disinformation}
\bibinfo{author}{Zannettou, S.} \emph{et~al.}
\newblock \bibinfo{title}{Disinformation warfare: Understanding state-sponsored trolls on twitter and their influence on the web}.
\newblock \emph{\bibinfo{journal}{Companion Proceedings of The 2019 World Wide Web Conference}} \bibinfo{pages}{218--226} (\bibinfo{year}{2019}).

\bibitem{bail2020assessing}
\bibinfo{author}{Bail, C.~A.} \emph{et~al.}
\newblock \bibinfo{title}{Assessing the russian internet research agency’s impact on the political attitudes and behaviors of american twitter users in late 2017}.
\newblock \emph{\bibinfo{journal}{Proceedings of the national academy of sciences}} \textbf{\bibinfo{volume}{117}}, \bibinfo{pages}{243--250} (\bibinfo{year}{2020}).

\bibitem{grinberg2019fake}
\bibinfo{author}{Grinberg, N.}, \bibinfo{author}{Joseph, K.}, \bibinfo{author}{Friedland, L.}, \bibinfo{author}{Swire-Thompson, B.} \& \bibinfo{author}{Lazer, D.}
\newblock \bibinfo{title}{Fake news on twitter during the 2016 us presidential election}.
\newblock \emph{\bibinfo{journal}{Science}} \textbf{\bibinfo{volume}{363}}, \bibinfo{pages}{374--378} (\bibinfo{year}{2019}).

\bibitem{eady2023exposure}
\bibinfo{author}{Eady, G.} \emph{et~al.}
\newblock \bibinfo{title}{Exposure to the russian internet research agency foreign influence campaign on twitter in the 2016 us election and its relationship to attitudes and voting behavior}.
\newblock \emph{\bibinfo{journal}{Nature Communications}} \textbf{\bibinfo{volume}{14}}, \bibinfo{pages}{62} (\bibinfo{year}{2023}).

\bibitem{bursztein2018quantifying}
\bibinfo{author}{Bursztein, E.} \& \bibinfo{author}{Marzuoli, A.}
\newblock \bibinfo{title}{Quantifying the impact of the twitter fake accounts purge-a technical analysis}.
\newblock \bibinfo{howpublished}{\url{https://elie.net/blog/web/quantifying-the-impact-of-the-twitter-fake-accounts-purge-a-technical-analysis/}}.
\newblock \bibinfo{note}{Accessed: 2023-05-25}.

\bibitem{roth2018twitter}
\bibinfo{author}{Roth, Y.} \& \bibinfo{author}{Harvey, D.}
\newblock \bibinfo{title}{How twitter is fighting spam and malicious automation}.
\newblock \bibinfo{howpublished}{\url{https://blog.twitter.com/en_us/topics/company/2018/how-twitter-is-fighting-spam-and-malicious-automation}}.
\newblock \bibinfo{note}{Accessed: 2023-05-25}.

\bibitem{bovet2019influence}
\bibinfo{author}{Bovet, A.} \& \bibinfo{author}{Makse, H.~A.}
\newblock \bibinfo{title}{Influence of fake news in twitter during the 2016 us presidential election}.
\newblock \emph{\bibinfo{journal}{Nature communications}} \textbf{\bibinfo{volume}{10}}, \bibinfo{pages}{7} (\bibinfo{year}{2019}).

\bibitem{flamino2023political}
\bibinfo{author}{Flamino, J.} \emph{et~al.}
\newblock \bibinfo{title}{Political polarization of news media and influencers on twitter in the 2016 and 2020 us presidential elections}.
\newblock \emph{\bibinfo{journal}{Nature Human Behaviour}} \bibinfo{pages}{1--13} (\bibinfo{year}{2023}).

\bibitem{hodges1958significance}
\bibinfo{author}{Hodges, J.~L.}
\newblock \bibinfo{title}{The significance probability of the smirnov two-sample test}.
\newblock \emph{\bibinfo{journal}{Arkiv f{\"o}r Matematik}} \textbf{\bibinfo{volume}{3}}, \bibinfo{pages}{469--486} (\bibinfo{year}{1958}).

\bibitem{mann1947test}
\bibinfo{author}{Mann, H.~B.} \& \bibinfo{author}{Whitney, D.~R.}
\newblock \bibinfo{title}{On a test of whether one of two random variables is stochastically larger than the other}.
\newblock \emph{\bibinfo{journal}{The annals of mathematical statistics}} \bibinfo{pages}{50--60} (\bibinfo{year}{1947}).

\bibitem{linvill2020troll}
\bibinfo{author}{Linvill, D.~L.} \& \bibinfo{author}{Warren, P.~L.}
\newblock \bibinfo{title}{Troll factories: Manufacturing specialized disinformation on twitter}.
\newblock \emph{\bibinfo{journal}{Political Communication}} \textbf{\bibinfo{volume}{37}}, \bibinfo{pages}{447--467} (\bibinfo{year}{2020}).

\bibitem{spirtes2000causation}
\bibinfo{author}{Spirtes, P.}, \bibinfo{author}{Glymour, C.} \& \bibinfo{author}{Scheines, R.}
\newblock \bibinfo{title}{Causation, prediction and search (2{\textordfeminine})} (\bibinfo{year}{2000}).

\bibitem{runge2012escaping}
\bibinfo{author}{Runge, J.}, \bibinfo{author}{Heitzig, J.}, \bibinfo{author}{Petoukhov, V.} \& \bibinfo{author}{Kurths, J.}
\newblock \bibinfo{title}{Escaping the curse of dimensionality in estimating multivariate transfer entropy}.
\newblock \emph{\bibinfo{journal}{Physical review letters}} \textbf{\bibinfo{volume}{108}}, \bibinfo{pages}{258701} (\bibinfo{year}{2012}).

\bibitem{runge2019detecting}
\bibinfo{author}{Runge, J.}, \bibinfo{author}{Nowack, P.}, \bibinfo{author}{Kretschmer, M.}, \bibinfo{author}{Flaxman, S.} \& \bibinfo{author}{Sejdinovic, D.}
\newblock \bibinfo{title}{Detecting and quantifying causal associations in large nonlinear time series datasets}.
\newblock \emph{\bibinfo{journal}{Science advances}} \textbf{\bibinfo{volume}{5}}, \bibinfo{pages}{eaau4996} (\bibinfo{year}{2019}).

\bibitem{runge2015identifying}
\bibinfo{author}{Runge, J.} \emph{et~al.}
\newblock \bibinfo{title}{Identifying causal gateways and mediators in complex spatio-temporal systems}.
\newblock \emph{\bibinfo{journal}{Nature communications}} \textbf{\bibinfo{volume}{6}}, \bibinfo{pages}{8502} (\bibinfo{year}{2015}).

\bibitem{cleveland1990stl}
\bibinfo{author}{Cleveland, R.~B.}, \bibinfo{author}{Cleveland, W.~S.}, \bibinfo{author}{McRae, J.~E.} \& \bibinfo{author}{Terpenning, I.}
\newblock \bibinfo{title}{Stl: A seasonal-trend decomposition}.
\newblock \emph{\bibinfo{journal}{J. Off. Stat}} \textbf{\bibinfo{volume}{6}}, \bibinfo{pages}{3--73} (\bibinfo{year}{1990}).

\bibitem{golovchenko2020cross}
\bibinfo{author}{Golovchenko, Y.}, \bibinfo{author}{Buntain, C.}, \bibinfo{author}{Eady, G.}, \bibinfo{author}{Brown, M.~A.} \& \bibinfo{author}{Tucker, J.~A.}
\newblock \bibinfo{title}{Cross-platform state propaganda: Russian trolls on twitter and youtube during the 2016 us presidential election}.
\newblock \emph{\bibinfo{journal}{The International Journal of Press/Politics}} \textbf{\bibinfo{volume}{25}}, \bibinfo{pages}{357--389} (\bibinfo{year}{2020}).

\bibitem{linvill2019russians}
\bibinfo{author}{Linvill, D.~L.}, \bibinfo{author}{Boatwright, B.~C.}, \bibinfo{author}{Grant, W.~J.} \& \bibinfo{author}{Warren, P.~L.}
\newblock \bibinfo{title}{“the russians are hacking my brain!” investigating russia's internet research agency twitter tactics during the 2016 united states presidential campaign}.
\newblock \emph{\bibinfo{journal}{Computers in Human Behavior}} \textbf{\bibinfo{volume}{99}}, \bibinfo{pages}{292--300} (\bibinfo{year}{2019}).

\end{thebibliography}
%% if required, the content of .bbl file can be included here once bbl is generated
%%\input sn-article.bbl

\clearpage
\listoffigures

\clearpage
\listoftables

\end{document}